\documentclass[12pt]{elsarticle}
\usepackage{amsmath,amssymb}
\usepackage{bm}
\usepackage{subfigure}

\journal{Journal of Sound and Vibration}

\begin{document}
\begin{frontmatter}

\title{A Chebyshev-Tau spectral method for normal modes of underwater sound propagation with a layered marine environment}

\author{Houwang Tu}
\author{Yongxian Wang\corref{cor1}}
\ead{yxwang@nudt.edu.cn}
\author{Qiang Lan}
\author{Wei Liu}
\author{Wenbin Xiao}
\author{Shuqing Ma}

\cortext[cor1]{corresponding author}
\address{College of Meteorology and Oceanography, National University of Defense Technology, Changsha, China}

\begin{abstract}
The normal mode model is one of the most popular approaches for solving underwater sound propagation problems. Among other methods, the finite difference method is widely used in classic normal mode programs. In many recent studies, the spectral method has been used for discretization. It is generally more accurate than the finite difference method. However, the spectral method requires that the variables to be solved are continuous in space, and the traditional spectral method is powerless for a layered marine environment. A Chebyshev-Tau spectral method based on domain decomposition is applied to the construction of underwater acoustic normal modes in this paper. In this method, the differential equation is projected onto spectral space from the original physical space with the help of an orthogonal basis of Chebyshev polynomials. A complex matrix eigenvalue / eigenvector problem is thus formed, from which the solution of horizontal wavenumbers and modal functions can be solved. The validity of the acoustic field calculation is tested in comparison with classic programs. The results of analysis and tests show that compared with the classic finite difference method, the proposed Chebyshev-Tau spectral method has the advantage of high computational accuracy. In addition, in terms of running time, our method is faster than the Legendre-Galerkin spectral method.
\end{abstract}

\begin{graphicalabstract}
\centering
\includegraphics[width=\linewidth] {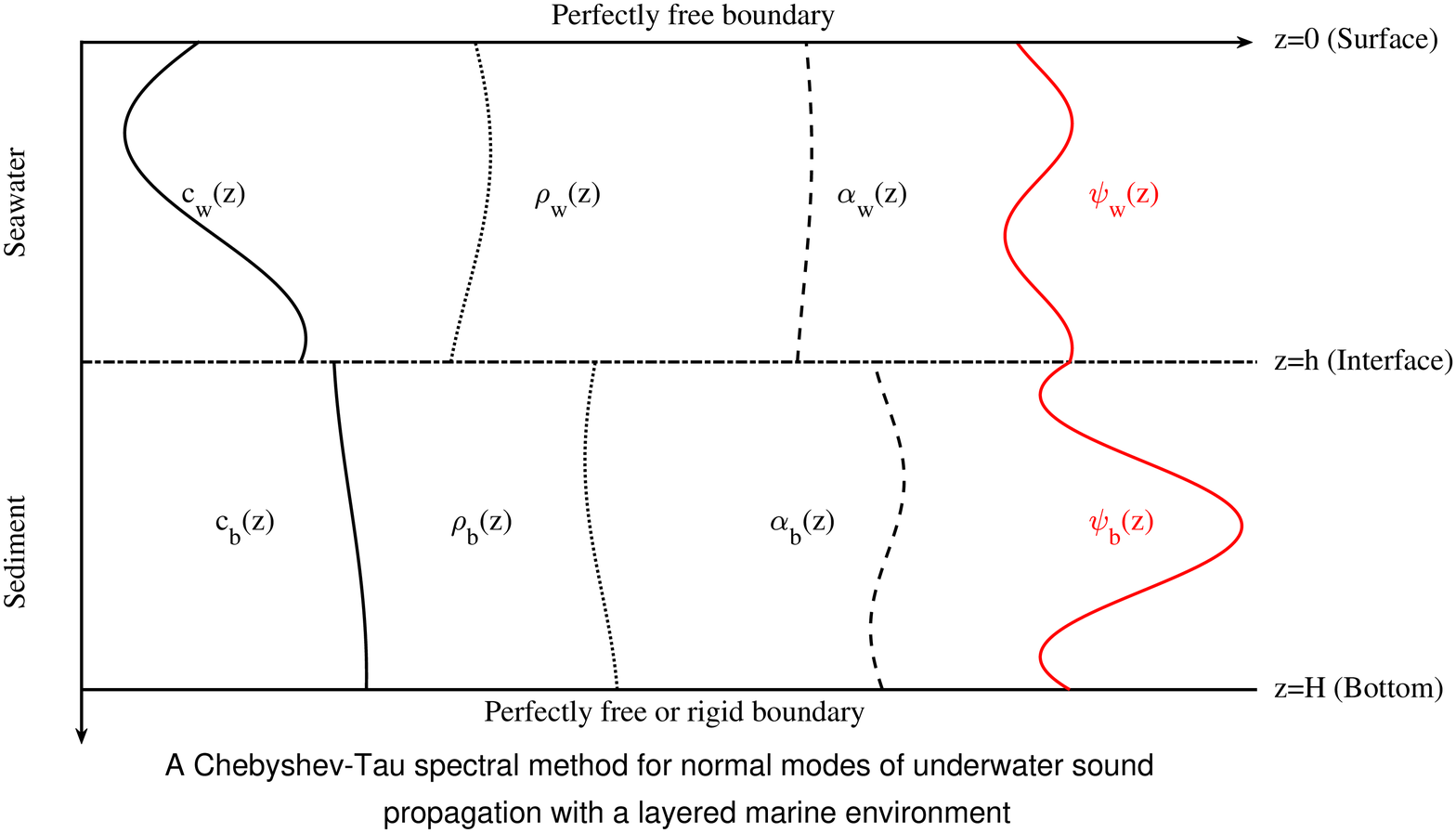}
{\scriptsize
A Chebyshev-Tau spectral method for normal modes of underwater sound propagation with a layered marine environment}
\end{graphicalabstract}

\begin{highlights}
\item Research highlight 1
: This paper proposes a Chebyshev-Tau spectral method for calculating acoustic propagation in a layered marine environment.
\item Research highlight 2
: The accuracy of the proposed method is higher than that of the classic finite difference method.
\item Research highlight 3
: Regarding the running time, our Chebyshev-Tau spectral method is far faster than the Legendre-Galerkin spectral method and slightly slower than the classic finite difference method.
\end{highlights}

\begin{keyword}
Chebyshev-Tau method \sep
spectral method \sep
normal modes \sep
underwater sound propagation \sep
computational ocean acoustics
\end{keyword}
\end{frontmatter}

\section{Introduction}

Sound waves are the main means of transmitting information remotely underwater. The study of the laws of underwater acoustic propagation helps investigators develop and utilize the ocean efficiently. The propagation of underwater sound waves can be modeled by the basic wave equation. Solving the wave equation directly is very difficult, but by approximating the wave equation from different perspectives, many mature underwater acoustic propagation models have gradually been developed. The depth separated wave equation for normal modes is common in underwater acoustics. One of the earliest papers was published in 1948 by Pekeris \cite{Pekeris1948}, who developed a theory for a simple two-layer model (ocean and sediment) with a constant sound speed in each layer. Progress in the development of normal mode methods is presented in an excellent summary by Williams \cite{Williams1970} published in 1970. Today, there are many models available that are based on normal modes \cite{Stickler1975,Porter1984}. In terms of numerical discretization, the finite difference method is the most common method of discretizing the normal mode model, although some studies have used other methods, such as the finite element method, to discretize the normal mode model \cite{Isakson2007,Vendhana2010,Ray2018}. As one of the widely used programs of the normal mode model in the acoustics community, Kraken \cite{Porter2001} was developed based on the finite difference method and has the advantages of robustness, accuracy and high efficiency.

In engineering numerical simulations, the spectral method is a numerical discrete method with higher accuracy than the finite difference method \cite{Gottlieb1977}. Since the 1970s, with the discovery of the fast Fourier transform and the development of high-performance computing technology, the shortcoming of the large amount of calculations of the spectral methods is gradually being overcome. Due to the advantages of high accuracy and fast convergence \cite{Canuto2006}, the spectral method has been widely used in the research studies of many disciplines, such as meteorology \cite{Zheng1989}, oceanography \cite{Canuto2006}, and physical and chemical measurements \cite{Muravskaya2019}. A number of researchers have tried to introduce spectral methods in the study of acoustic problems. Adamou et al. \cite{Adamou2004} proposed an efficient, accurate, and flexible numerical scheme based on spectral methods to determine the dispersion curves and displacement/stress profiles for modes in elastic guiding structures, possibly curved, layered, damped, inhomogeneous, or anisotropic. Quintanilla et al. \cite{Quintanilla2015} attempted to model guided elastic waves in generally anisotropic media using a spectral collocation method. Colbrook et al. \cite{Colbrook2019} presented a new approach for solving acoustic scattering problems. Wise et al. \cite{Wise2019} presented an arbitrary acoustic source and sensor distributions in a Fourier collocation method. Wang et al. \cite{Wang2006} presented an improved expansion scheme for the acoustical wave propagator. Evans \cite{Evans2016} proposed a Legendre-Galerkin technique for differential eigenvalue problems with complex and discontinuous coefficients in underwater acoustics. This method requires that the lower boundary be the pressure release boundary, and the computational speed is slow. Subsequently, Evans et al. \cite{Evans2018} studied the Legendre-Galerkin spectral method for constructing atmospheric acoustic normal modes. Tu et al. \cite{Tu2019a} showed normal mode and parabolic equation models using the Chebyshev spectral method to process a single layer of a body of water with constant density and no attenuation.

In this article, we introduce a Chebyshev spectral method that can address the discontinuity of the sound speed, density and attenuation profiles based on the normal mode model. This method projects the original underwater acoustic propagation problem onto a finite-dimensional subspace with Chebyshev polynomials as the basis functions \cite{Boyd2001}. After solving the problem in the spectral space, the expansion coefficients are inversely transformed back to the real physical space. The Tau method is used in this study to perform the spectral expansion of the boundary conditions and explicitly apply the constraints of the boundary conditions to the solution of the spectral coefficients. In the proposed method, it is necessary to consider the discontinuity of the parameters at the interface of the water column and bottom sediment; thus, domain decomposition is used in the depth direction to address this issue \cite{Min2005}.

\section{Normal mode model in a layered marine environment}
\begin{figure}[htb]
\centering
\includegraphics[width=\linewidth] {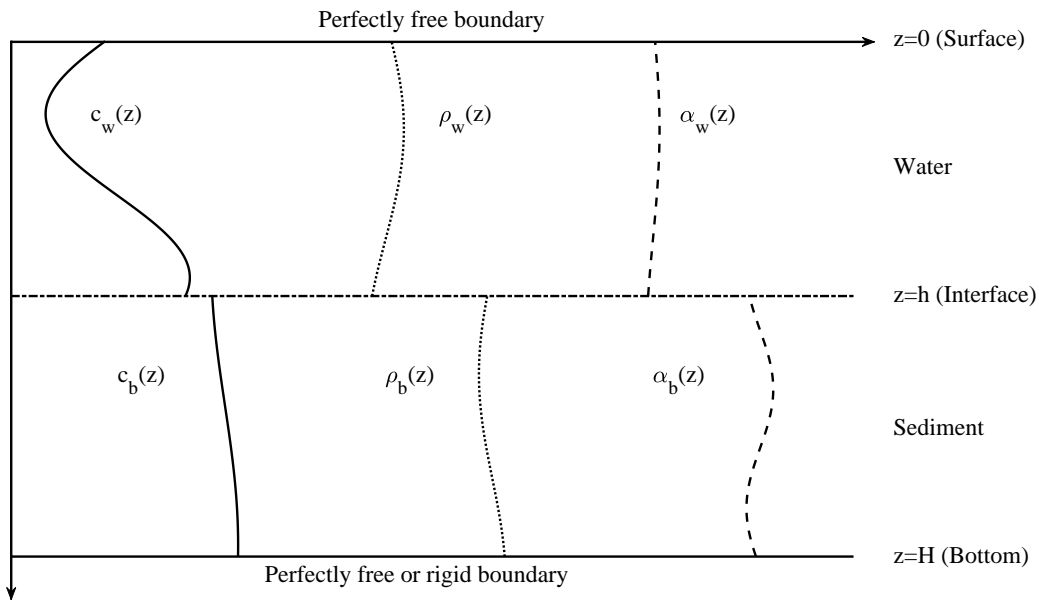}
\caption{Sound speed, density and attenuation profiles of the layered marine environment.}
\label{fig:1}
\end{figure}

Considering a cylindrical coordinate system where the sound source is a simple harmonic point source and the marine environment is a cylindrically symmetric two-dimensional sound propagation case, the Helmholtz equation can be written as
\begin{equation}
\label{eq:1}
    \frac{1}{r}\frac{\partial}{\partial r}\left(
        r\frac{\partial p}{\partial r}
    \right) + 
    \rho (z) \frac{\partial}{\partial z}\left(
        \frac{1}{\rho (z)}\frac{\partial p}{\partial z}
    \right) +
    \frac{\omega ^2}{c(z)^2}p=0
\end{equation}
where $\omega=2\pi f$ is the angular frequency and $f$ is the frequency of the sound source. By using the technique of the separation of variables, the time-independent sound pressure $p(r,z)$ can be decomposed into
\begin{equation}
\label{eq:2}
    p(r,z)=\psi(z)R(r)
\end{equation}
where $R(r)$ satisfies the Hankel function, and $\psi (z)$ satisfies the following modal equation:
\begin{equation}
\label{eq:3}
    \rho(z)\frac{\mathrm{d}}{\mathrm {d}z}\left(
        \frac{1}{\rho(z)}\frac{\mathrm{d}\psi(z)}{\mathrm {d}z}
    \right) + k^2\psi(z) = k_r^2 \psi(z)
\end{equation}
where $k_r^2$ is a separation constant and $k_r$ is the horizontal wavenumber.
The function $\rho(z)$ is the density, $k=(1+i\eta\alpha)\omega / c(z)$ is the complex wavenumber, $\alpha$ is the attenuation in dB$/  \lambda$ ($\lambda$ is the wavelength of the sound source), and $\eta=(40\pi \log_{10}{e})^{-1}$. In this paper, a two-layer media problem consisting of a layer of water ($z\in [0,h]$) and a layer of sedimentation ($z\in [h,H]$) is considered. As shown in Fig.~\ref{fig:1}, the marine environmental parameters in Eq.~\eqref{eq:3} are defined in the water and bottom separately, and they can be discontinuous at the interface $z=h$:
\begin{equation}
\begin{split}
\label{eq:4}
    c(z) &= \begin{cases}
        c_w(z),&0 \leq z \leq h\\
        c_b(z),&h\leq z \leq H
    \end{cases},
    \cr
    \rho(z) &= \begin{cases}
        \rho_w(z),&0 \leq z \leq h\\
        \rho_b(z),&h \leq z\leq H
    \end{cases},
    \cr
    \alpha(z) &= \begin{cases}
        \alpha_w(z),&0 \leq z \leq h\\
        \alpha_b(z),&h\leq z \leq H
    \end{cases}
\end{split}
\end{equation}

The boundary conditions at the ends of the interval $0 \leq z \leq H$ and the interface conditions at $z=h$, where $h\leq H$, are imposed. The upper boundary condition is
\begin{equation}
\label{eq:5}
    \psi(z=0)=0
\end{equation}
The lower boundary condition is:
\begin{equation}
\label{eq:6}
    \psi(z=H)=0
\end{equation}
or
\begin{equation}
\label{eq:7}
    \frac{\mathrm{d}\psi(z=H)} {\mathrm{d}z}=0
\end{equation}
which correspond to a perfectly free surface at $z=0$ and a perfectly free or rigid bottom at $z=H$. At the interface, the sound pressure and the normal particle velocity of sound pressure must be continuous; these conditions are imposed by allowing for a unique value $\psi$ at the interface:
\begin{equation}
\label{eq:8}
    \psi(z=h^{-} )=\psi(z=h^{+})
\end{equation}
\begin{equation}
\label{eq:9}
    \frac{1}{\rho(z=h^{-})} \frac{\mathrm{d}\psi(z=h^{-} )}{\mathrm {d}z}= \frac{1}{\rho(z=h^{+})} \frac{\mathrm{d}\psi(z=h^{+} )}{\mathrm {d}z}
\end{equation}
where the superscripts $-$ and $+$ indicate limits from above and below, respectively. Eq.~\eqref{eq:3} has a set of solutions $(k_{r,m}, \psi_m), m=1, 2, \dots$ when supplemented by the above boundary conditions, where $k_{r,m}$ and $\psi_m$ are also called the $m$-th horizontal wavenumber and eigenmode, respectively. The eigenmodes of Eq.~\eqref{eq:3} are arbitrary to a non-zero scaling constant. We assume that the modes are scaled (normalized) so that
\begin{equation}
\label{eq:10}
    \int_{0}^{H} \frac{{[\psi_m(z)}]^2}{\rho(z)}\mathrm {d} z=1,
    \quad m = 1, 2, \dots, M
\end{equation}
where $m$ is the order of modes and $M$ is the number of modes. Finally, the fundamental solution of the 2D Helmholtz equation can be approximated by
\begin{equation}
\label{eq:11}
p(r,z) \approx \frac{i}{4\rho(z_s)}\sum_{m=1}^{M}\psi_m(z_s)\psi_m(z)H_0^{(1)}(k_{r,m}r)
\end{equation}
where $z_s$ is the depth of the sound source, and $H_0^{(1)}(\cdot)$ is the Hankel function, which corresponds to $R(r)$.

To demonstrate the sound field results, the transmission loss (TL) of the sound pressure is defined as follows:
\begin{equation}
\label{eq:12}
    \text{TL}=-20\log_{10}\left(\frac{|p|}{|p_0|}\right)
\end{equation}
The unit of TL is the decibel (dB), where $p_0$ is the sound pressure at a distance of 1 m from the sound source. In an actual sound field display, TL is usually used.

\section{Chebyshev-Tau spectral method for the normal mode model}
\label{sec:3}

\subsection{Chebyshev spectral method}

The spectral method is based on using finite-order function expansion and truncation to approximate the original function. The continuous smooth function $u(x)$ is expanded according to a set of basis functions in a weighted sum form. This set of basis functions, also known as trial functions, constitutes a set of orthogonal bases in a continuous function space. If the basis functions are Chebyshev orthogonal polynomials, the Chebyshev spectral method is formed. In this case, the trial function $T_k (x)$ is
\begin{equation}
\label{eq:13}
T_k(x)=\cos{(k\cos^{-1}(x))},
\quad k = 0, 1, 2, \cdots
\end{equation}
Any smooth differentiable function $u(x)$ defined in the interval $[-1,1]$ can be expanded according to this set of basis functions as follows:
\begin{equation}
\label{eq:14}
u(x)=\sum_{k=0}^{\infty}\hat{u}_k T_k(x)
\end{equation}
We use the hat symbol to represent the Chebyshev spectral expansion coefficient of the corresponding function. For example, $\hat{u}_k$ and $\widehat{(uv)}_k$ represent the $k$-th spectral expansion coefficients of functions $u$ and $uv$, respectively. The expansion coefficient $\hat{u}_k$ is
\begin{equation}
\label{eq:15}
    \hat{u}_k = \frac{2}{\pi c_k} \int_{-1}^{1} \frac{u(x)T_k(x)}{\sqrt{1-x^2}} \mathrm{d} x,
    \quad c_k=\begin{cases}
        2,  & k=0\\
        1,  & k>0
    \end{cases}
\end{equation}
Eqs.~\eqref{eq:14} and \eqref{eq:15} are called the inverse Chebyshev transform and forward Chebyshev transform, respectively.

When numerically calculating the integral in Eq.~\eqref{eq:15}, it is necessary to introduce numerical discretization for the variable $x$ in the interval $[-1,1]$. Considering the convenience of the processing boundary conditions and the accuracy of the numerical calculations, the Gauss-Lobatto nodes are often used \cite{Canuto2006}:
\begin{equation}
\label{eq:16}
    x_j=\cos{\left(\frac{j\pi}{N}\right)},
    \quad j=0,1,2,\dots,N
\end{equation}
where $N$ is the number of discretization points, which is equal to the truncation order in the spectral method. In this case, the expansion coefficient $\hat{u}_k$ of Eq.~\eqref{eq:15} can be approximately calculated as
\begin{equation}
\label{eq:17}
\begin{split}
    \hat{u}_k \approx \frac{1}{d_k}\sum_{j=0}^{N}u(x_j)T_k(x_j)\omega_j,
    \quad k=0,1,2,\dots,N \\
    \omega_j=\begin{cases}
    \frac{\pi}{2N},\quad j=0,N  \\
    \frac{\pi}{N}, \quad \text{otherwise}
 \end{cases},
    d_k=\begin{cases}
    \pi,\quad k=0,N  \\
    \frac{\pi}{2},\quad \text{otherwise}
 \end{cases}
\end{split}
\end{equation}
and the discretization version of Eq.~\eqref{eq:14} is thus
\begin{equation}
\label{eq:18}
u(x) \approx \sum_{k=0}^{N}\hat{u}_k T_k(x)
\end{equation}

Similar to expanding the function $u(x)$ in Eq.~\eqref{eq:14}, the first derivative function of $u(x)$ can also be expanded approximately as
\begin{equation}
\label{eq:19}
    u'(x)\approx\sum_{k=0}^{N}\hat{u}'_k T_k(x)
\end{equation}
The expansion coefficients satisfy the following relationship:
\begin{equation}
\label{eq:20}
    \hat{u}'_k \approx \frac{2}{c_k}
        \sum_{
        \substack{j=k+1,\\ 
            j+k=\text{odd}
            }}^{N} j \hat{u}_j
\end{equation}
where $c_k$ is defined in Eq.~\eqref{eq:15}. Equivalently, we can also express Eq.~\eqref{eq:20} in the form of a matrix-vector product
\begin{equation}
\label{eq:21}
    \hat{\mathbf{u}}' \approx \mathbf{D} \hat{\mathbf{u}}
\end{equation}
where $\hat{\mathbf{u}}' = [\hat{u}'_0, \dots, \hat{u}'_N]^T$, $\hat{\mathbf{u}} = [\hat{u}_0, \dots, \hat{u}_N]^T$, and $\mathbf{D}$ is a square matrix of order $(N+1)$. The superscript $T$ denotes the transpose of a vector.

The spectral coefficients of the product of two known functions can also be represented by the respective spectral coefficients \cite{Canuto2006}:
\begin{equation}
\label{eq:22}
    \widehat{(uv)}_k \approx 
        \frac{1}{2} \sum_{m+n=k}^{N} \hat{u}_m\hat{v}_n +
        \frac{1}{2} \sum_{|m-n|=k}^{N} \hat{u}_m\hat{v}_n
\end{equation}
its equivalent vector form is
\begin{equation}
\label{eq:23}
    \widehat{\mathbf{(uv)}} \approx \mathbf{C}_v \hat{\mathbf{u}}
\end{equation}
where $\widehat{\mathbf{(uv)}} = [\widehat{(uv)}_0, \dots, \widehat{(uv)}_N]^T$ and $\mathbf{C}_v$ is a $(N+1)$-order square matrix, whose elements depend on the values of $\hat{v}_n$.

To normalize each mode by using Eq.~\eqref{eq:10}, we must evaluate the integration of a known function $u(x)$ in interval $[-1,1]$. Here, we give only the following formula briefly \cite{Mason2003}:
\begin{equation}
 \label{eq:24}
    \int_{-1}^{1}u(x) \text{d} x \approx -2\sum_{
    \substack{n=0,\\
    n=\text{even}}
    }^{N}\frac{\hat u_n}{n^2-1}
 \end{equation}

\subsection{Discrete normal mode model using the Chebyshev-Tau spectral method}

The construction of the Chebyshev polynomial spanning starts with a choice of the maximum order $N$ of the polynomials, which depends on the number of modes needed, as discussed below. A single set of Chebyshev polynomials cannot span both the water column and bottom sediment since it will not have the required derivative discontinuity at the water-sediment interface. It is advisable to use the domain decomposition method of Min and Gottleib \cite{Min2005}. The domain decomposition takes the form
\begin{equation}
\label{eq:25}
\psi(z)= \begin{cases}
    \psi_w(z)\approx\sum_{k=0}^{N}\hat{\psi}_{w_k}T_k(x_w),\quad x_w=-\frac{2}{h}z+1,&
       0\leq z\leq h \\
    \psi_b(z)\approx\sum_{k=0}^{N}\hat{\psi}_{b_k}T_k(x_b),\quad
    x_b=\frac{2}{h-H}z+\frac{H+h}{H-h},&
      h\leq z\leq H
 \end{cases}
\end{equation}
Both $\psi_w(z)$ and $\psi_b(z)$ satisfy Eq.~\eqref{eq:3} on $[0,h]$ and $[h, H]$, respectively.

To obtain the discretization equation of Eq.~\eqref{eq:3}, as the first step, we introduce the coordinate transformation of $z \in I \mapsto x \in [-1,1]$. Here, the interval is $I = [0,h]$ for the water column layer and $I = [h,H]$ for the bottom sediment layer.
By applying the coordinate transformation to Eq.~\eqref{eq:3}, we obtain
\begin{equation}
\label{eq:26}
\frac{4}{|I|^2}\rho(x)\frac{\mathrm{d}}{\mathrm{d}x}\left(\frac{1}{\rho(x)}\frac{\mathrm{d}\psi(x)}{\mathrm {d}x}\right) +k^2\psi(x) = k_r^2 \psi(x),
\quad
x \in [-1, 1]
\end{equation}
Here, we use the fact that ${\mathrm{d}x} / {\mathrm {d}z}=-2 / |I|$, and $|I|$ denotes the length of the interval $I$.
In the next step, we discretize the problem into a system of linear equations. Constructing linear equations by the spectral method requires the residual error to be orthogonal to a set of finite dimensional bases. Classically, the Tau, Galerkin, collocation and pseudospectral methods are commonly used for this purpose. We discuss the Tau method in detail in this paper, which is essentially a method of mean weighted residuals. When applied to solve Eq.~\eqref{eq:3}, the Chebyshev-Tau method aims to obtain the solution of the weak form of Eq.~\eqref{eq:3} (see Eq.~(3.3.15) of \cite{Canuto2006} for further details):
\begin{equation}
\label{eq:27}
\begin{split}
\int_{-1}^{1}\left[\frac{4}{|I|^2}\rho(x)\frac{\mathrm{d}}{\mathrm{d}x}\left(\frac{1}{\rho(x)}\frac{\mathrm{d}\psi(x)}{\mathrm {d}x}\right) +k^2\psi(x)-k_r^2 \psi(x) \right]\frac{T_k(x)}{\sqrt{1-x^2}}\mathrm{d}x=0\\
x\in (-1,1), \quad k=0,1,\dots,N-2
\end{split}
\end{equation}
The orthogonality of Chebyshev polynomials $\{T_k(x)\}$ can lead to a system of $(N-1)$ linear equations, and the two endpoints of the interval $I$ are constrained by boundary conditions. For brevity in the following text, we introduce some temporary functions $v(x), w(x), s(x)$ on $[-1, 1]$ defined as
\begin{equation}
\label{eq:28}
    v(x)= \frac{1}{\rho(x)}, 
    \quad
    w(x)= v(x) \psi'(x), 
    \quad
    s(x)= \rho(x) w'(x)
\end{equation}
By following these conventions, we apply Eqs.~\eqref{eq:21} and \eqref{eq:23} to Eq.~\eqref{eq:28} twice, and we can obtain
\begin{equation}
\label{eq:29}
    \hat{\mathbf{w}}= -\frac{2}{|I|}\mathbf{C}_v(\mathbf{D}\hat{\bm{\psi})},\quad
    \hat{\mathbf{s}}= -\frac{2}{|I|}\mathbf{C}_{\rho}(\mathbf{D}\hat{\mathbf{w}})
    = \frac{4}{|I|^2}\mathbf{C}_{\rho}(\mathbf{D}(\mathbf{C}_v(\mathbf{D}\hat{\bm{\psi}})))
\end{equation}
As a result, the matrix form of Eq.~\eqref{eq:3} in the Chebyshev-Tau method can be finally written as
\begin{equation}
\label{eq:30}
\left(\frac{4}{|I|^2}\mathbf{C}_{\rho}\mathbf{D}\mathbf{C}_v\mathbf{D}+\mathbf{C}_{k^2}\right)\hat{\bm{\psi}}=k_r^2\hat{\bm{\psi}}
\end{equation}
Applying this result to the water column with $|I| = h$ and the bottom sediment with $|I|=H-h$, we can obtain the following:
\begin{equation}
\label{eq:31}
\begin{aligned}
    \mathbf{A}
    \hat{\bm{\psi}}_w
    &= k_r^2
    \hat{\bm{\psi}}_w,
    &
    \mathbf{A} 
    &=
    \frac{4}{h^2}\mathbf{C}_{\rho}\mathbf{D}\mathbf{C}_v\mathbf{D}+\mathbf{C}_{k^2},
\\
    \mathbf{B}
    \hat{\bm{\psi}}_b
    &= k_r^2
    \hat{\bm{\psi}}_b,
    &
    \mathbf{B} 
    &= 
    \frac{4}{(H-h)^2}\mathbf{C}_{\rho}\mathbf{D}\mathbf{C}_v\mathbf{D}+\mathbf{C}_{k^2}
\end{aligned}
\end{equation}

Both matrices $\mathbf{A}$ and $\mathbf{B}$ are square matrices of order $(N+1)$, and we split matrix $\mathbf{A}$, for example, into four blocks:
\begin{equation}
\label{eq:32}
\mathbf{A}=\left[\begin{array}{cc}
\mathbf{A}_{11}&\mathbf{A}_{12}\\
\mathbf{A}_{21}&\mathbf{A}_{22}\\
\end{array}\right]
\end{equation}
where block matrices $\mathbf{A}_{11}$ and $\mathbf{A}_{22}$ are square matrices of order $(N-1)$ and 2, respectively. In a similar way, we can also obtain four block matrices of $\mathbf{B}$ to be $\mathbf{B}_{11}$, $\mathbf{B}_{12}$, $\mathbf{B}_{21}$ and $\mathbf{B}_{22}$.

As the weak form of Eq.\eqref{eq:3} shows, the Tau method does not require each trial function to satisfy the boundary conditions, which are directly enforced by the equations of the system. Instead, the boundary conditions and interface conditions Eqs.~\eqref{eq:5} to \eqref{eq:9} are expanded with Eq.~\eqref{eq:17} and transformed into linear equations about $\hat{\bm{\psi}}_w$ and $\hat{\bm{\psi}}_b$. In the question we studied, the boundary conditions and the interface conditions yield four linear equations that determine four of the coefficients of the two highest-order polynomials. For convenience of description, we introduce the intermediate row vectors

\begin{equation*}
\mathbf{s}=[s_0,s_1,s_2,\dots,s_N],\quad \mathbf{t}=[t_0,t_1,t_2,\dots,t_N]   
\end{equation*}
where $s_k=T_k(-1)=(-1)^k$, $t_k=T_k(+1)=1$. Furthermore, let

\begin{equation*}
\mathbf{p}=\frac{1}{\rho_w(h)}\mathbf{s}\mathbf{D},\quad \mathbf{q}=\frac{1}{\rho_b(h)}\mathbf{t}\mathbf{D}.
\end{equation*}
Thus, the boundary conditions and interface conditions of Eqs.~\eqref{eq:5} to \eqref{eq:9} can be transformed and expressed as

\begin{align}
\label{eq:33}
    &\mathbf{t}\hat{\bm{\psi}}_w=0  \\
\label{eq:34}
    &\mathbf{s}\hat{\bm{\psi}}_b=0  \\
\label{eq:35}
    &\mathbf{s}\mathbf{D}\hat{\bm{\psi}}_b=0\\
\label{eq:36}
    & \mathbf{s}\hat{\bm{\psi}}_w-\mathbf{t}\hat{\bm{\psi}}_b=0 \\
\label{eq:37}
    &\mathbf{p}\hat{\bm{\psi}}_w-\mathbf{q}\hat{\bm{\psi}}_b=0
\end{align}

We divide the four row vectors into blocks, and we name the row vectors consisting of the first $(N-1)$ elements as $\mathbf{s}_1$, $\mathbf{t}_1$, $\mathbf{p}_1$, and $\mathbf{q}_1$ and the row vectors consisting of the last two elements as $\mathbf{s}_2$, $\mathbf{t}_2$, $\mathbf{p}_2$, and $\mathbf{q}_2$, respectively. Similarly, let the first $(N-1)$ elements of the column vectors $\hat{\bm{\psi}}_{w}$ and $\hat{\bm{\psi}}_{b}$ be $\hat{\bm{\psi}}_{w1}$ and $\hat{\bm{\psi}}_{b1}$, respectively. Since the interface conditions must consider both the water and the bottom sediment, Eq.~\eqref{eq:34} is taken as the lower boundary condition as an example. The matrix form of Eq.~\eqref{eq:31} becomes
\begin{equation}
\label{eq:38}
\left[\begin{array}{cc|cc}
\mathbf{A}_{11}&\mathbf{0}&\mathbf{A}_{12}&\mathbf{0}\\
\mathbf{0}&\mathbf{B}_{11}&\mathbf{0}&\mathbf{B}_{12}\\
\hline
\mathbf{t}_1&\mathbf{0}&\mathbf{t}_2&\mathbf{0}\\
\mathbf{s}_1&-\mathbf{t}_1&\mathbf{s}_2&-\mathbf{t}_2\\
\mathbf{p}_1&-\mathbf{q}_1&\mathbf{p}_2&-\mathbf{q}_2\\
\mathbf{0}&\mathbf{s}_1&\mathbf{0}&\mathbf{s}_2
\end{array}\right]
\left[\begin{array}{c}
\hat{\bm{\psi}}_{w1}\\
\hat{\bm{\psi}}_{b1}\\
\hline
\hat{\bm{\psi}}_{w,N-1}\\
\hat{\bm{\psi}}_{w,N}\\
\hat{\bm{\psi}}_{b,N-1}\\
\hat{\bm{\psi}}_{b,N}\\
\end{array}
\right]=k_r^2\left[
\begin{array}{c}
\hat{\bm{\psi}}_{w1}\\
\hat{\bm{\psi}}_{b1}\\
\hline
0\\
0\\
0\\
0\\
\end{array}\right]
\end{equation}
According to the block structure of the matrix and/or vectors in the above formula, Eq.~\eqref{eq:38} can be abbreviated as
\begin{equation}
\label{eq:39}
\left[
\begin{array}{cc}
\mathbf{L}_{11}&\mathbf{L}_{12}\\
\mathbf{L}_{21}&\mathbf{L}_{22}\\
\end{array}
\right]\left[
\begin{array}{c}
\hat{\bm{\psi}}_1\\
\hat{\bm{\psi}}_2
\end{array}
\right]=k_r^2\left[
\begin{array}{c}
\hat{\bm{\psi}}_1\\
\mathbf{0}
\end{array}\right]
\end{equation}
where $\mathbf{L}_{11}$ is a square matrix of order $(2N-2)$, and $\mathbf{L}_{22}$ is a square matrix of order 4. The $(2N-2)$-dimensional column vector $\hat{\bm{\psi}}_1$ is obtained by stacking the $(N-1)$-dimensional column vectors $\hat{\bm{\psi}}_{w1}$ and $\hat{\bm{\psi}}_{b1}$ and the 4-dimensional column vector $\hat{\bm{\psi}}_2=[\hat{\bm{\psi}}_{w,N-1},\hat{\bm{\psi}}_{w,N},\hat{\bm{\psi}}_{b,N-1},\hat{\bm{\psi}}_{b,N}]^T$. The remaining matrices satisfy the relevant compatibility conditions. The resulting system of Eq.~\eqref{eq:39} can be solved by the following two steps:
\begin{equation}
\label{eq:40}
\hat{\bm{\psi}}_2=-\mathbf{L}_{22}^{-1}\mathbf{L}_{21}\hat{\bm{\psi}}_1,\quad(\mathbf{L}_{11}-\mathbf{L}_{12}\mathbf{L}_{22}^{-1}\mathbf{L}_{21})\hat{\bm{\psi}}_1=k_r^2\hat{\bm{\psi}}_1
\end{equation}
Therefore, it is necessary only to solve the $(2N-2)$-order algebraic eigenvalue problem in Eq.~\eqref{eq:40}. For each set of eigenvalues / eigenvectors $(k_r^2,\hat{\bm{\psi}})$, a set of eigensolutions $(k_r,\bm{\psi})$ of Eq.~\eqref{eq:3} can be obtained by the discrete inverse Chebyshev transform Eq.~\eqref{eq:18}. In this process, it is necessary to inversely transform the eigenvectors $\hat{\bm{\psi}}_w$ and $\hat{\bm{\psi}}_b$ to the depth at the same resolution. The vectors $\bm{\psi}_w$ and $\bm{\psi}_b$ should be stacked into a single column vector to form $\bm{\psi}$, and each mode should be normalized by Eq.~\eqref{eq:10}. Finally, the sound pressure field is obtained by applying Eq.~\eqref{eq:11} to every mode chosen by the favorable phase speed interval. A systematic theoretical study of the convergence of the Chebyshev polynomials and Tau method can be found in Boyd \cite{Boyd2001}, and the asymptotic rate of convergence in the Chebyshev-Tau spectral method has been previously reported \cite{Coleman1976}; no more details are provided here.

\section{Test examples and analysis of results}

To verify the validity of the Chebyshev-Tau spectral method in solving the normal mode model, the following tests and analysis are performed through seven examples listed in Table \ref{tab1}. To make these examples closer to the real ocean, the attenuation of seawater in the marine environment is set to $\alpha_w=0.0$. For comparison, we introduce the widely used Kraken program based on finite difference discretization and the rimLG program \cite{Evans2016} based on the Legendre-Galerkin spectral method. To facilitate the description, the normal mode model program based on the Chebyshev-Tau spectral method developed by us is abbreviated as the NM-CT program. Since the Kraken program is based on Fortran language and the rimLG program is based on MATLAB language, we developed both a Fortran code version and a MATLAB code version of NM-CT. The numerical error of the results of the two versions is less than $10^{-14}$. The results listed in the following text retain only ten decimal places; thus, the results reported by the two implementations should be the same. It must be mentioned that the size of $N$ in NM-CT depends on the number of modes required for the final synthesized sound field, and the number of modes is usually related to factors such as depth and sound source frequency. Therefore, the choice of $N$ is empirical. We estimate the number of modes of the final synthesized sound field through the depth, sound source frequency, and modal phase velocity limits, similar to what Kraken did; the size of $N$ can usually be 2 to 10 times the number of modes. For a large number of measured profile data, the size of $N$ is recommended to be 6 to 20 times the number of modes.

\begin{figure}[htbp]
\centering
\subfigure[]{\includegraphics[width=4.2cm]{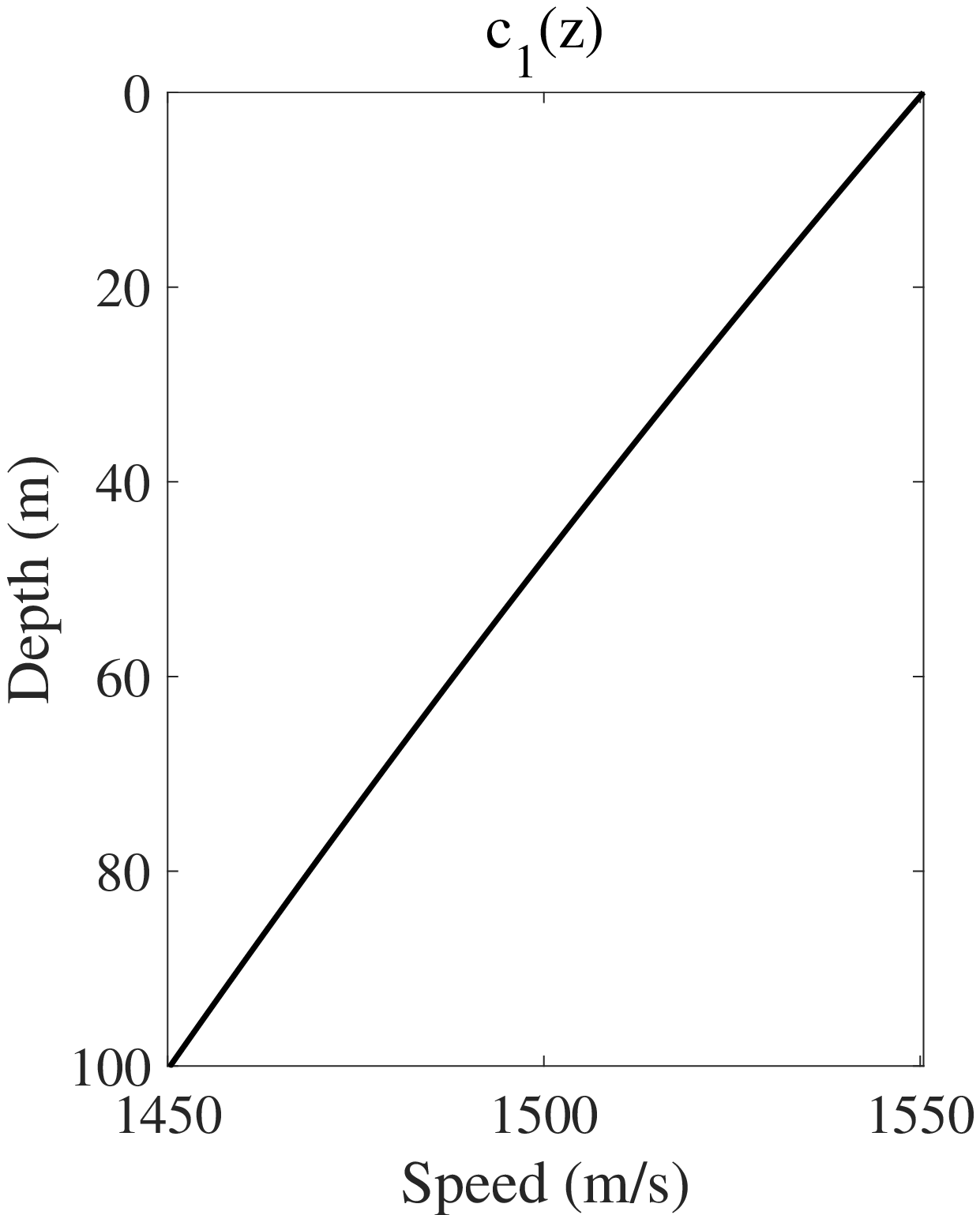}}
\subfigure[]{\includegraphics[width=4.2cm]{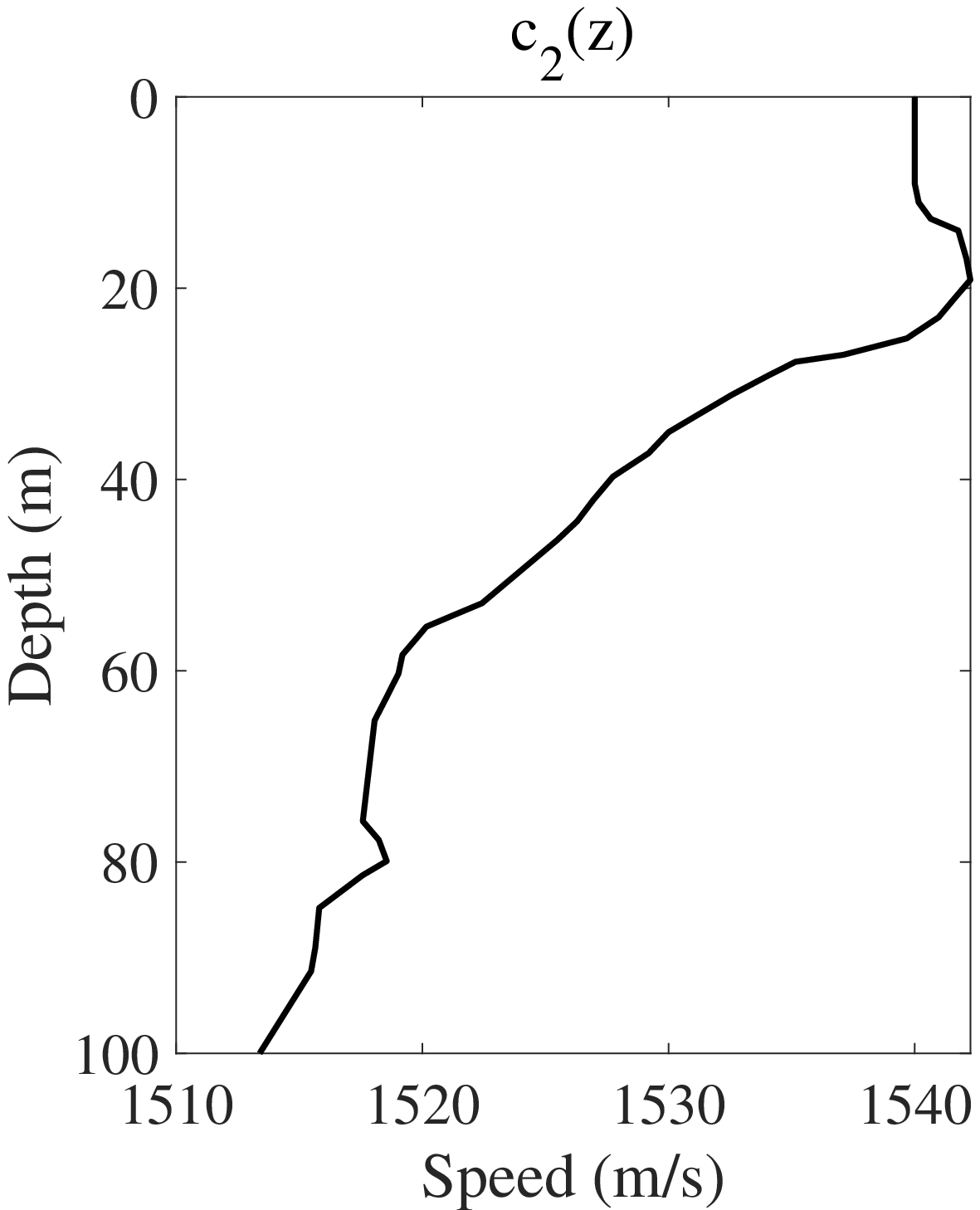}}
\subfigure[]{\includegraphics[width=4.2cm]{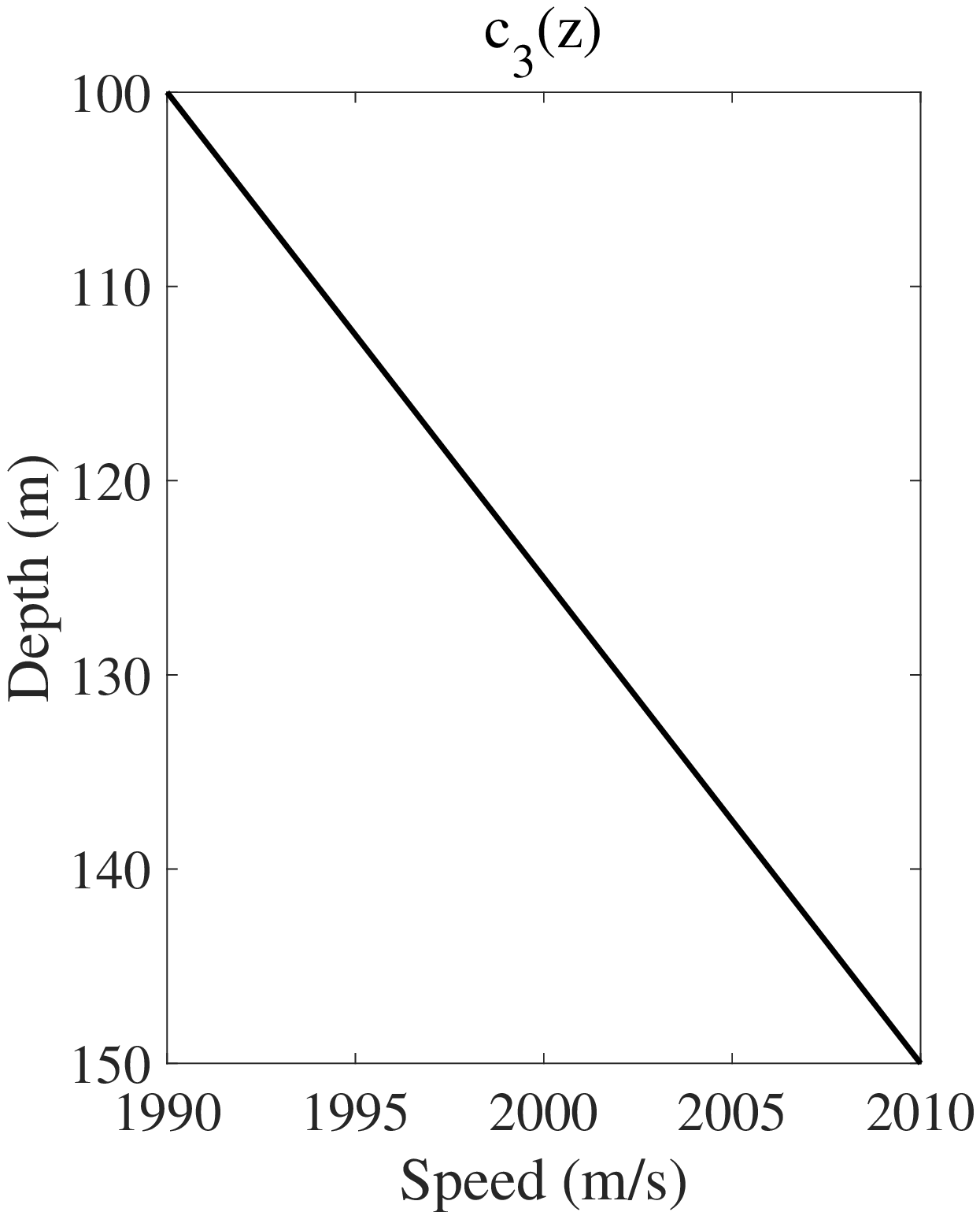}}
\subfigure[]{\includegraphics[width=4.2cm]{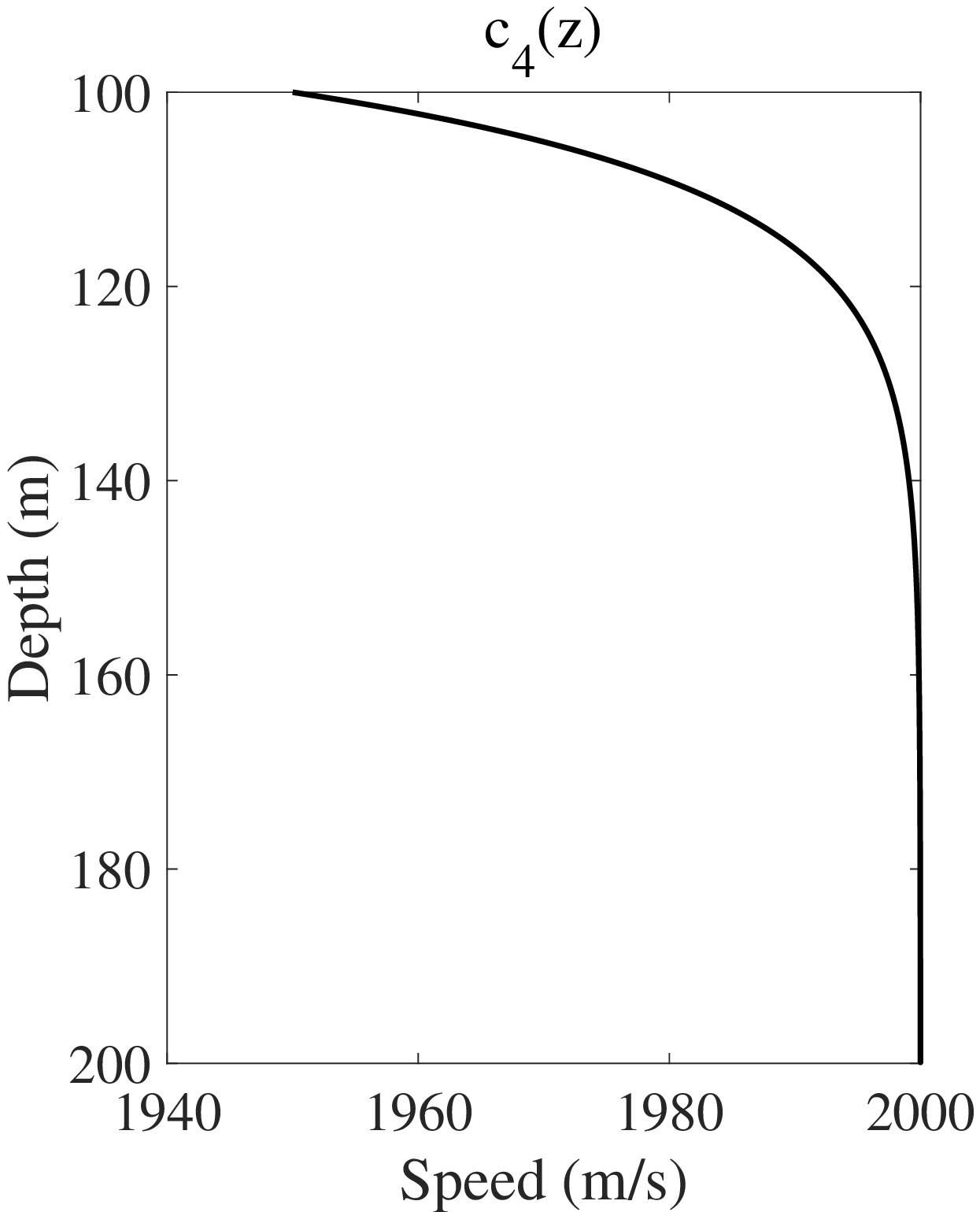}}
\subfigure[]{\includegraphics[width=4.2cm]{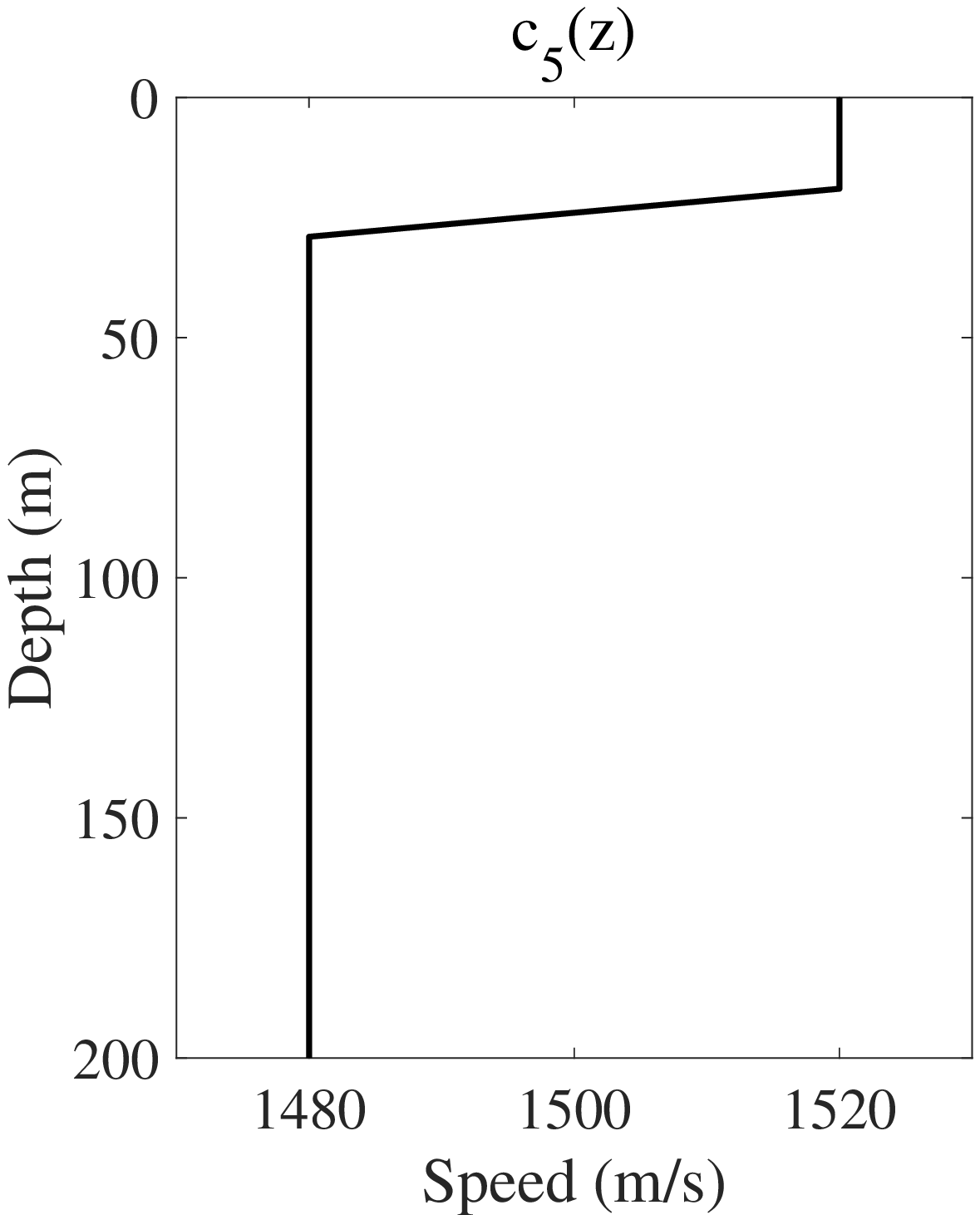}}
\subfigure[]{\includegraphics[width=4.2cm]{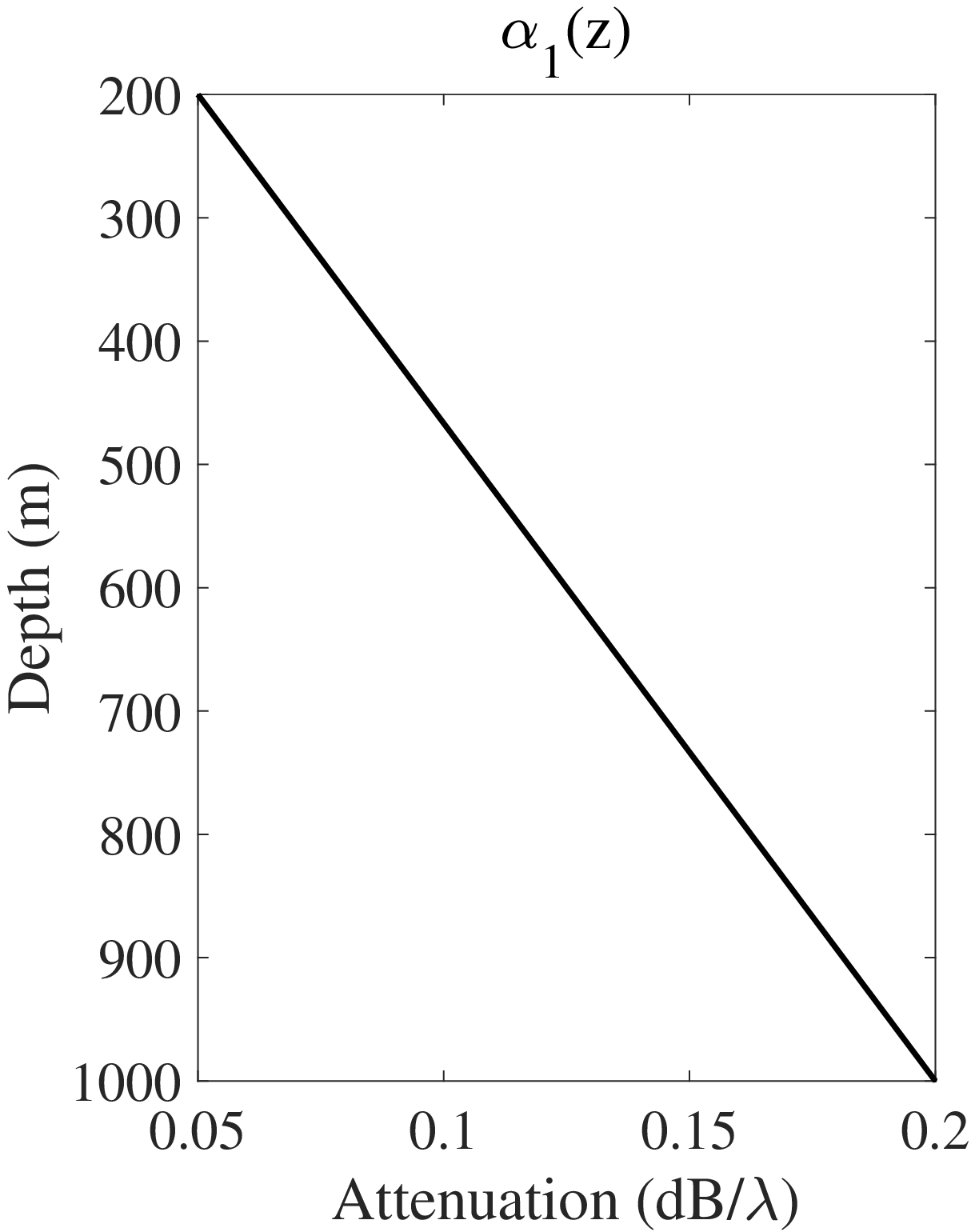}}
\caption{Acoustic profiles in the examples.}
\label{fig:2}
\end{figure}

The various profiles used for the numerical examples are
\begin{enumerate}
\item
The pseudolinear profile \cite{Finn2011} is shown in Fig.~\ref{fig:2}(a) and given by:
\begin{equation}
\label{eq:41}
c_1(z)=\frac{1}{\sqrt{az+b}}, a=5.94\times10^{-10} \text{s}^2/\text{m}^3, b = 4.16\times10^{-7} \text{s}^2/\text{m}^3
\end{equation}
\item
The surface duct profile \cite{Chowdhury2018} $c_2(z)$ is shown in Fig.~\ref{fig:2}(b).
\item
The linear sound speed profile in the bottom layer is shown in Fig.~\ref{fig:2}(c) and given by
\begin{equation}
\label{eq:42}
c_3(z)=0.4z+1950
\end{equation}
\item
The exponentially increasing sound speed profile in the bottom layer is shown in Fig.~\ref{fig:2}(d) and given by
\begin{equation}
\label{eq:43}
c_4(z)=2000-50\exp{\left(-\frac{z-100}{10}\right)}
\end{equation}
\item
The segmented sound speed profile in the water column is shown in Fig.~\ref{fig:2}(e) and given by
\begin{equation}
\label{eq:44}
c_5(z)=\begin{cases}
    1520,\quad 0 \le z \le 20  \\
    1600-4z,\quad 20 \le z \le 30  \\
    1480,\quad 30 \le z \le 200 
 \end{cases}
\end{equation}
\item
The other linear increasing sound speed profile in the bottom sediment is given by
\begin{equation}
\label{eq:45}
c_6(z)=0.5z+1500
\end{equation}
The plot of $c_6(z)$ is similar to that of $c_3(z)$ and is not shown in Fig.~\ref{fig:2}.
\item
The linear increasing attenuation profile in the bottom layer is shown in Fig.~\ref{fig:2}(f) and given by
\begin{equation}
\label{eq:46}
\alpha_1(z)=0.0001875z+0.0125
\end{equation}
\end{enumerate}

\begin{table}[htbp]
\footnotesize
\centering
\begin{tabular}{cccccccccc}
\hline
    {Number of}&
    ${\rho_w}$&
    ${c_w}$& 
    ${\alpha_w}$& 
    ${h}$& 
    ${\rho_b}$& 
    ${c_b}$& 
    ${\alpha_b}$ & 
    ${H}$ &
    {Bottom}
\\
    {Examples}        &
    {g/cm$^3$}        &
    {m/s}             &
    {dB/$\lambda$}    &
    {m}               &
    {g/cm$^3$}        &
    {m/s}             &
    {dB/$\lambda$}    &
    {m}               &
    {free/rigid}
\\
\hline
    {Example 1}& 
    {1.0} &
    {$c_1$} &
    {0.0} &
    {50}&
    {1.0}&
    {$c_1$}&
    {0.0}&
    {100}&
    {Eq.~\eqref{eq:7}}
\\
    {Example 2}& 
    {1.0} &
    {1500} &
    {0.0}&
    {50}&
    {1.5}&
    {1800}&
    {1.5}&
    {100}&
    {Eq.~\eqref{eq:6}}
\\
    {Example 3}& 
    {1.0} &
    {$c_2$}& 
    {0.0}&
    {100}&
    {2.0}&
    {$c_3$}&
    {0.0}&
    {150}&
    {Eq.~\eqref{eq:6}}
\\
    {Example 4}& 
    {1.0} &
    {$c_2$}& 
    {0.0}&
    {100}&
    {2.0}&
    {$c_3$}&
    {0.5}&
    {150}&
    {Eq.~\eqref{eq:6}}
\\
    {Example 5}& 
    {1.0} &
    {$c_1$}&
    {0.0}&
    {100}&
    {2.0}&
    {$c_4$}&
    {0.5}&
    {200}&
    {Eq.~\eqref{eq:7}}
\\
    {Example 6}& 
    {1.0} &
    {$c_1$}&
    {0.0}&
    {100}&
    {2.0}&
    {$c_4$}&
    {0.5}&
    {200}&
    {Eq.~\eqref{eq:6}}
\\
    {Example 7}& 
    {1.0} &
    {$c_5$}&
    {0.0}&
    {200}&
    {1.5}&
    {$c_6$}&
    {$\alpha_1$}&
    {1000}&
    {Eq.~\eqref{eq:6}}
\\
\hline
\end{tabular}
\caption{\label{tab1} List of examples.}
\end{table}

\subsection{Single-layer waveguide}

\begin{figure}[htbp]
\centering
\subfigure[]{\includegraphics[width=6.5cm]{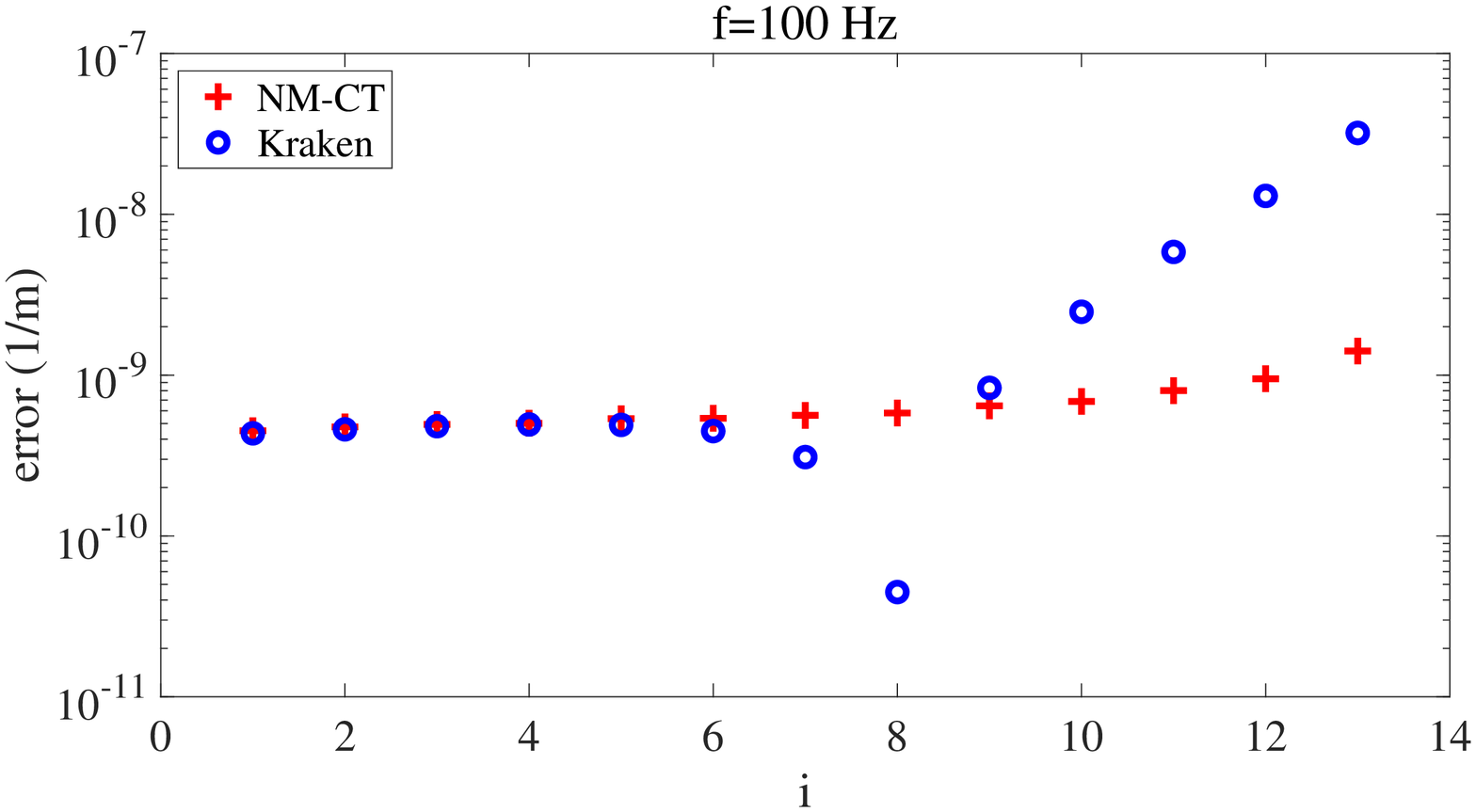}}
\subfigure[]{\includegraphics[width=6.5cm]{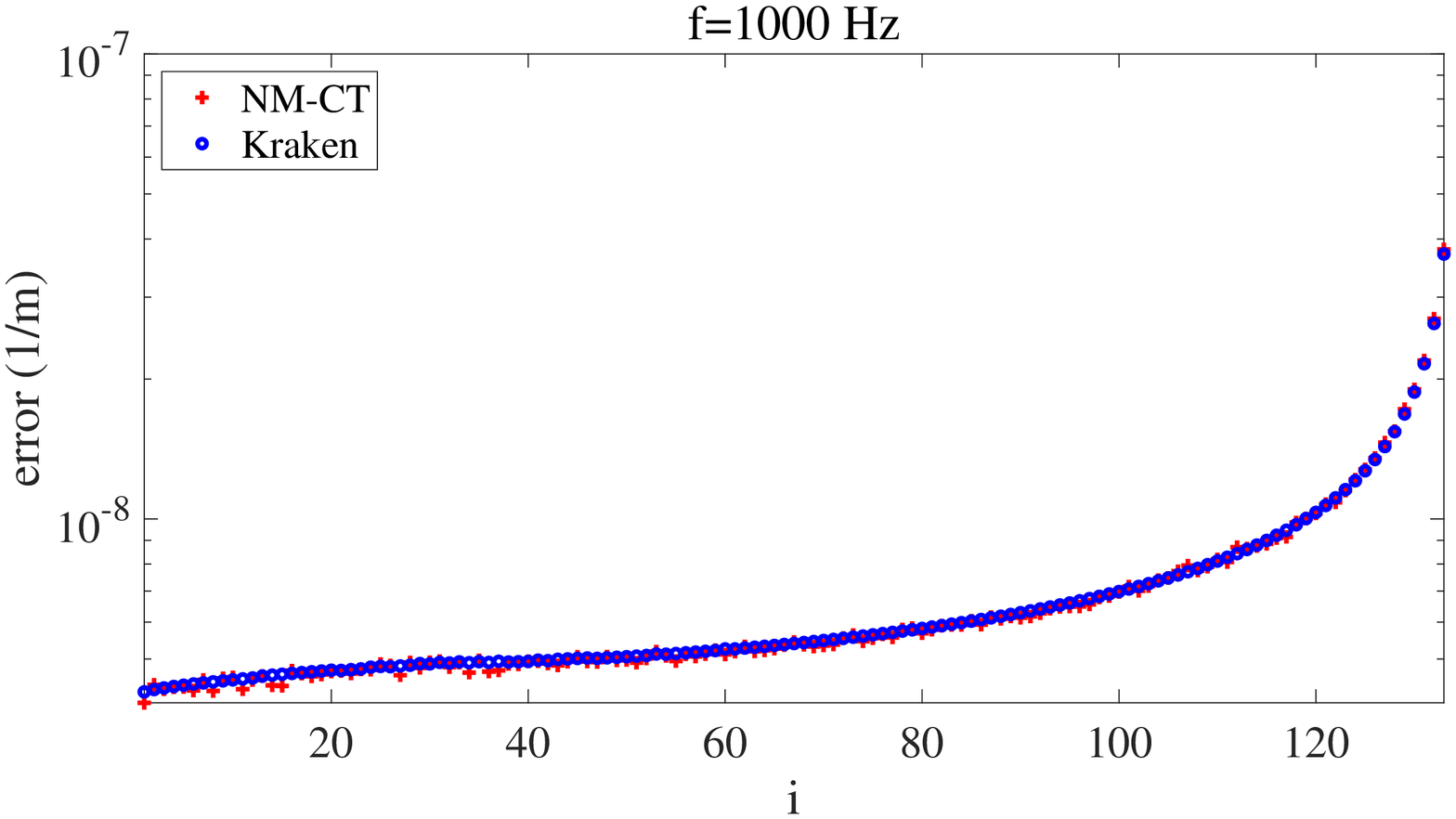}}
\subfigure[]{\includegraphics[width=13.4cm]{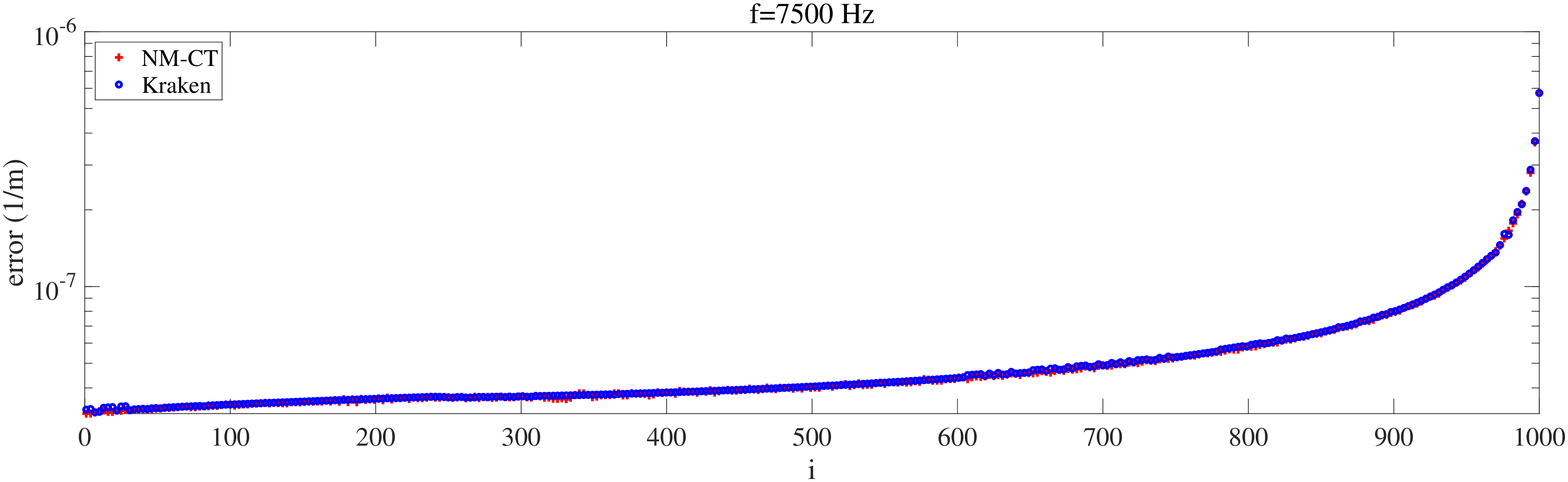}}
\caption{The errors of the wavenumbers calculated by the Kraken (small blue circles) and NM-CT (red plus sign) programs compared with the analytical solutions.}
\label{fig:3}
\end{figure}

Example 1 of Table \ref{tab1} is a single-layer waveguide problem. The density of seawater is constant at 1 g/cm$^3$, and there is no attenuation in the water. For this type of single-layer marine environment, the NM-CT program developed based on the Chebyshev-Tau spectral method is competent and needs only to set the acoustic parameters in the two layers as a continuous layer in the input file. In example 1, the pseudolinear sound speed waveguide has an analytical solution given by the Airy functions \cite{Finn2011}. The reliability of the NM-CT program is proven by comparison with analytical solutions. Three source frequencies, $f$ = 100 Hz, 1000 Hz and 7500 Hz, are considered. Because rimLG can handle only the perfectly free lower boundary, it does not appear in example 1. Table \ref{tab2} shows a few horizontal wavenumbers of example 1 calculated by the analytical solution, Kraken and NM-CT program for three frequencies. NM-CT uses a spectral truncation order of $N$=100 at 100 Hz, $N$=400 at 1000 Hz and $N$=1000 at 7500 Hz, and 400, 2000 and 6000 equidistant discrete points are used in the three frequencies for the Kraken program. It can be seen in the table that the wavenumbers calculated by the three programs are very similar; the Kraken result matches the analytical solutions with seven significant digits for the three frequencies. The NM-CT matches the analytical solutions with eight, seven and seven significant digits for $f$ = 100 Hz, 1000 Hz and 7500 Hz, respectively. Fig.~\ref{fig:3} shows the errors of the wavenumbers calculated by the two programs compared with the analytical solution. In the figure, it can be seen more intuitively that regardless of the frequency, the error of the wavenumber calculated by Kraken and NM-CT is approximately an order of magnitude in this example. However, Kraken uses far more discrete points in the vertical direction than NM-CT, which means that NM-CT needs fewer discrete points to achieve higher accuracy.
\begin{table}[htbp]
\scriptsize
\centering
\begin{tabular}{ccccc}
\hline
Freq.
&$m$
&Analytic 
&Kraken  
&NM-CT \\
\hline
$f$=100 Hz
& 1  & 0.428726 2353 & 0.428726 2349 & 0.428726 2349 \\
& 2  & 0.418720 6763 & 0.418720 6759 & 0.418720 6760 \\
& 3  & 0.411401 3864 & 0.411401 3859 & 0.411401 3859 \\
& 6  & 0.381781 7434 & 0.381781 7430 & 0.381781 7430 \\
& 7  & 0.366073 2255 & 0.366073 2252 & 0.366073 2250 \\
& 8  & 0.346775 4845 & 0.346775 4845 & 0.346775 4840 \\
& 11 & 0.258936 7165 & 0.258936 7224 & 0.258936 7158 \\
& 12 & 0.212957 4128 & 0.212957 4258 & 0.212957 4119 \\
& 13 & 0.147227 5253 & 0.147227 5574 & 0.147227 5241 \\
\hline
$f$=1000 Hz
&1   & 4.32256 5484 & 4.32256 5479 & 4.32256 5480 \\
&5   & 4.26193 0803 & 4.26193 0798 & 4.26193 0798 \\
&9   & 4.22231 1533 & 4.22231 1528 & 4.22231 1528 \\
&61  & 3.73510 9820 & 3.73510 9815 & 3.73510 9815 \\
&65  & 3.66898 6745 & 3.66898 6740 & 3.66898 6740 \\
&69  & 3.59716 8254 & 3.59716 8249 & 3.59716 8249 \\
&121 & 1.80461 2920 & 1.80461 2910 & 1.80461 2909 \\
&125 & 1.51324 5610 & 1.51324 5598 & 1.51324 5598 \\
&129 & 1.13647 1493 & 1.13647 1476 & 1.13647 1476 \\
\hline
$f$=7500 Hz
&1   &32.47270 963 & 32.47270 960 & 32.47270 960 \\
&32  &31.97126 325 & 31.97126 322 & 31.97126 322 \\
&63  &31.66388 669 & 31.66388 666 & 31.66388 666 \\
&156 &30.95233 922 & 30.95233 918 & 30.95233 918 \\
&218 &30.55369 016 & 30.55369 012 & 30.55369 012 \\
&311 &29.84310 998 & 29.84310 994 & 29.84310 994 \\
&404 &28.75361 660 & 28.75361 656 & 28.75361 656 \\
&528 &26.71692 999 & 26.71692 995 & 26.71692 995 \\
&621 &24.67372 146 & 24.67372 141 & 24.67372 141 \\
\hline
\end{tabular}
\caption{\label{tab2}Comparison of $k_{r,m}$ of example 1.}
\end{table}

\subsection{Two-layer waveguide}

Examples 2 to 7 of Table \ref{tab1} are layered waveguide problems. When the bottom sediment has attenuation, the modal wavenumbers are complex numbers. KrakenC is a version of Kraken that finds the wavenumbers in the complex plane, which computes the complex eigenvalues exactly. Therefore, we also tested the results of KrakenC under the same configuration as Kraken for all examples with attenuation.

\begin{figure}[htbp]
\centering
\subfigure[]{\includegraphics[width=6.5cm]{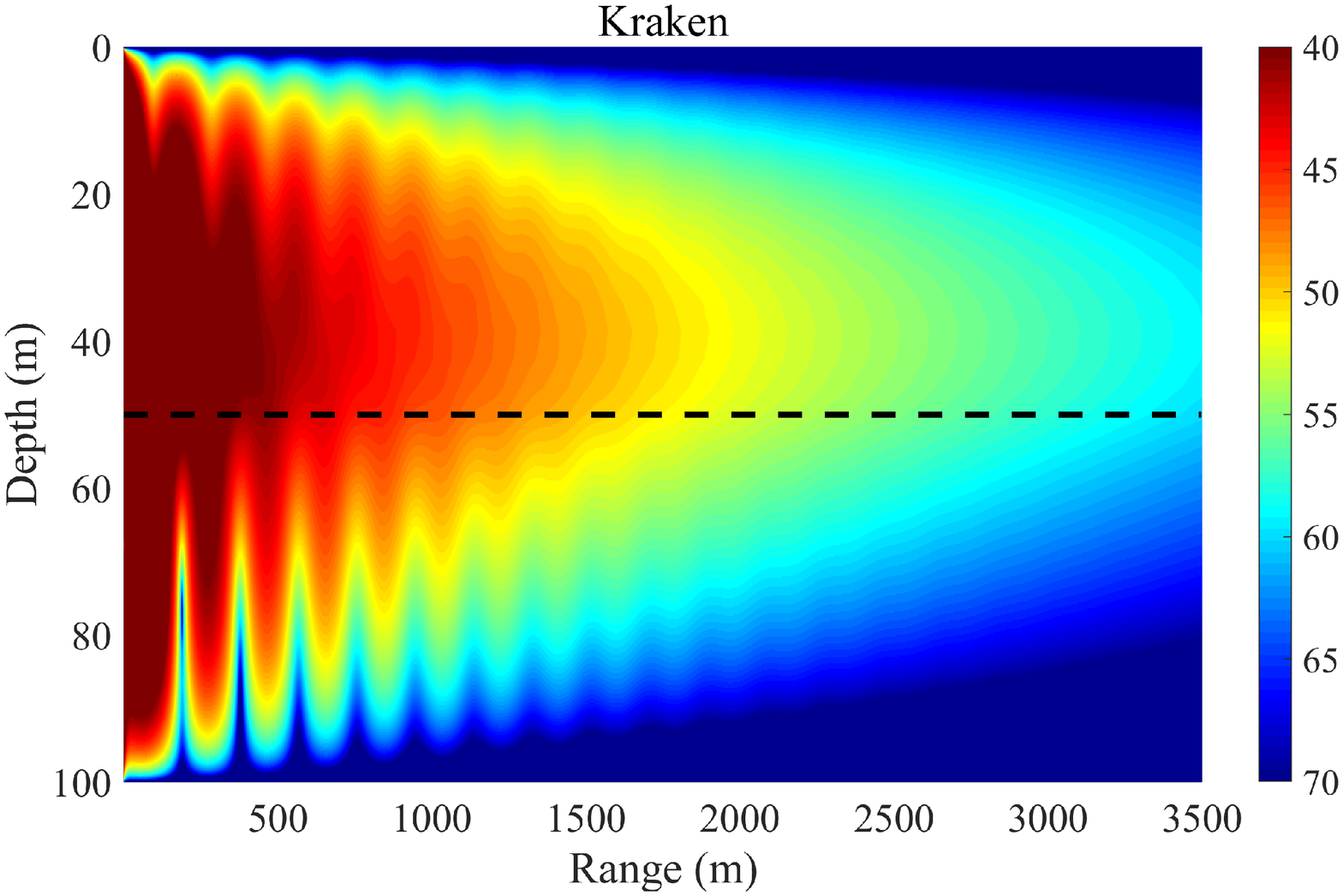}}
\subfigure[]{\includegraphics[width=6.5cm]{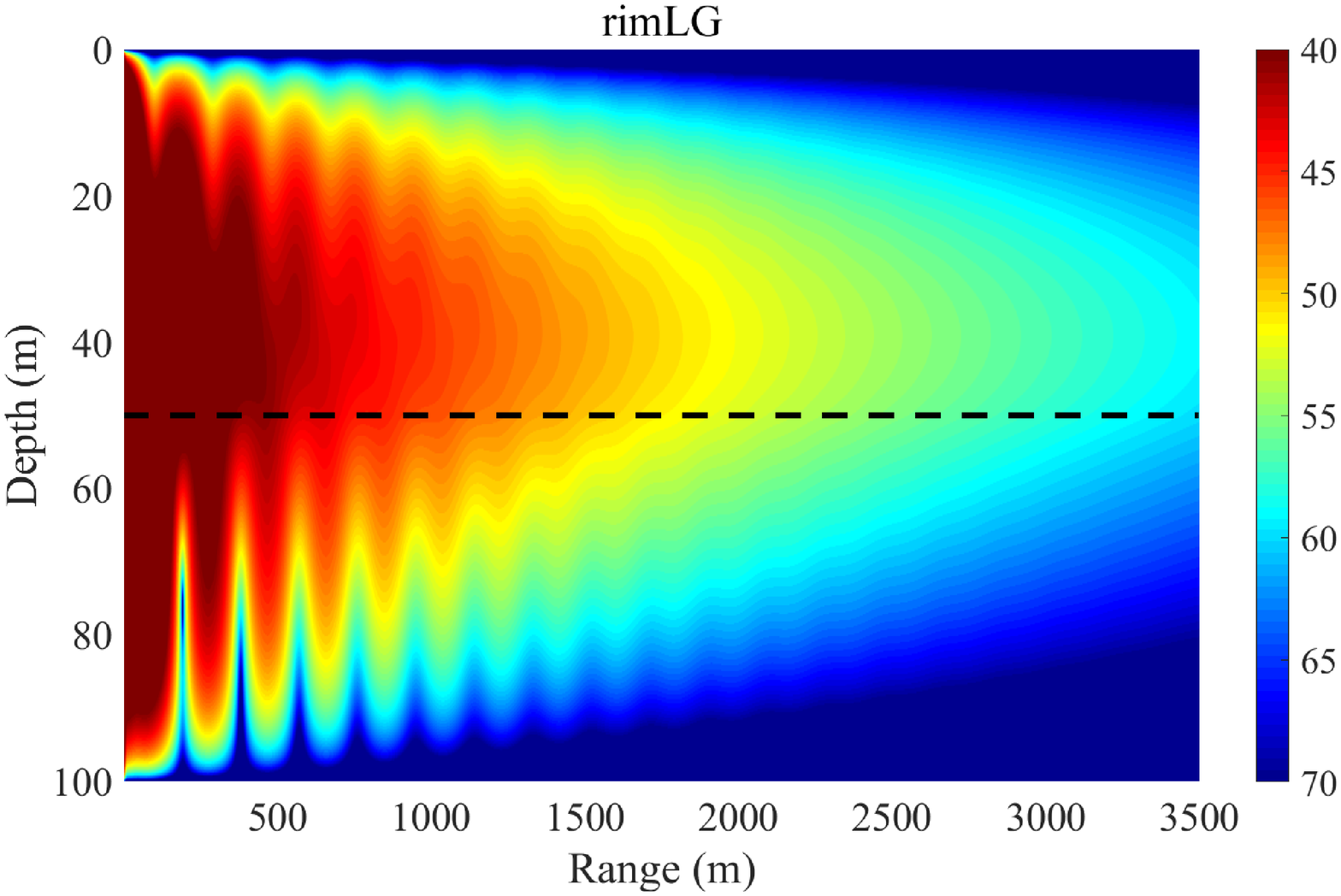}}
\subfigure[]{\includegraphics[width=6.5cm]{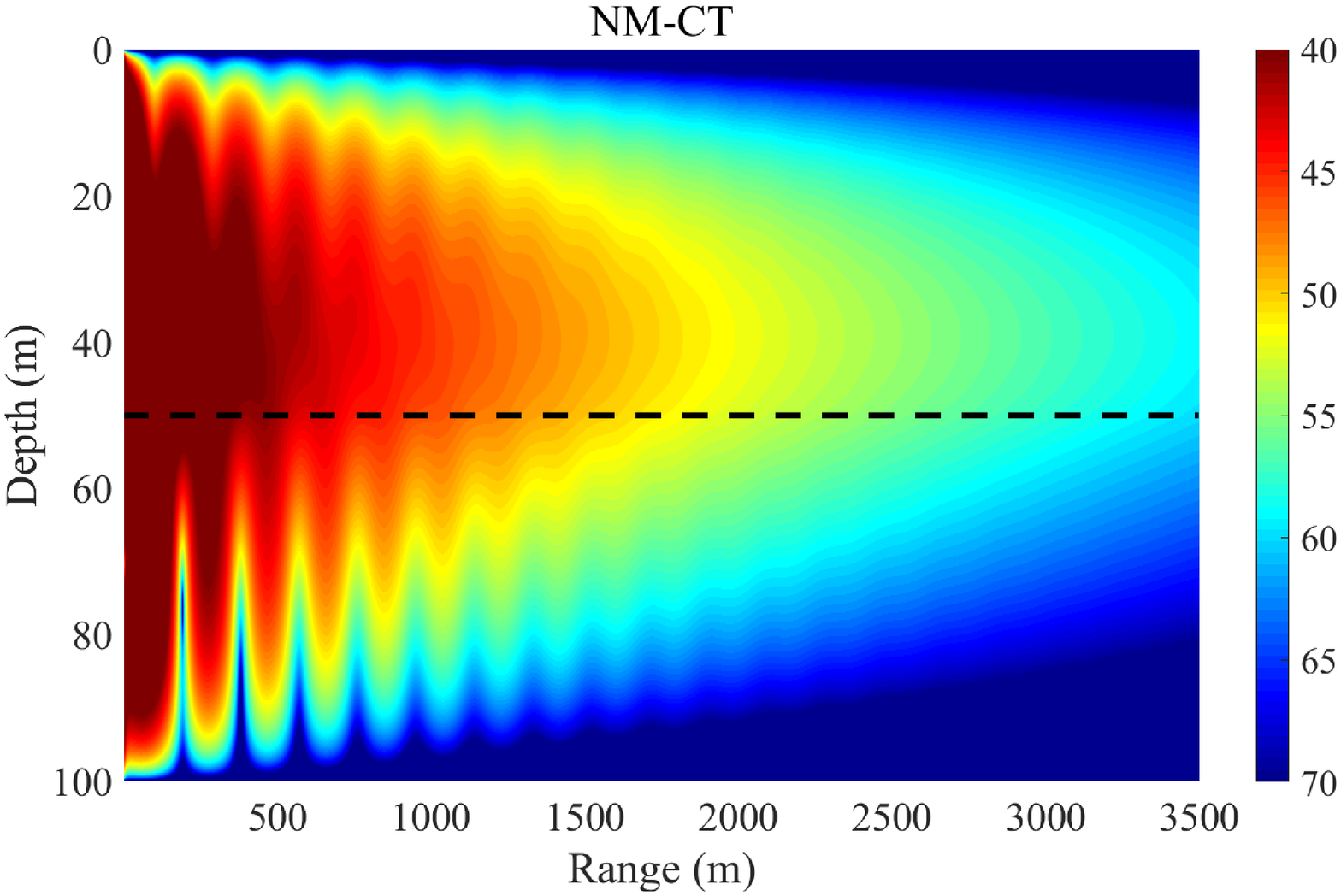}}
\subfigure[]{\includegraphics[width=6.5cm]{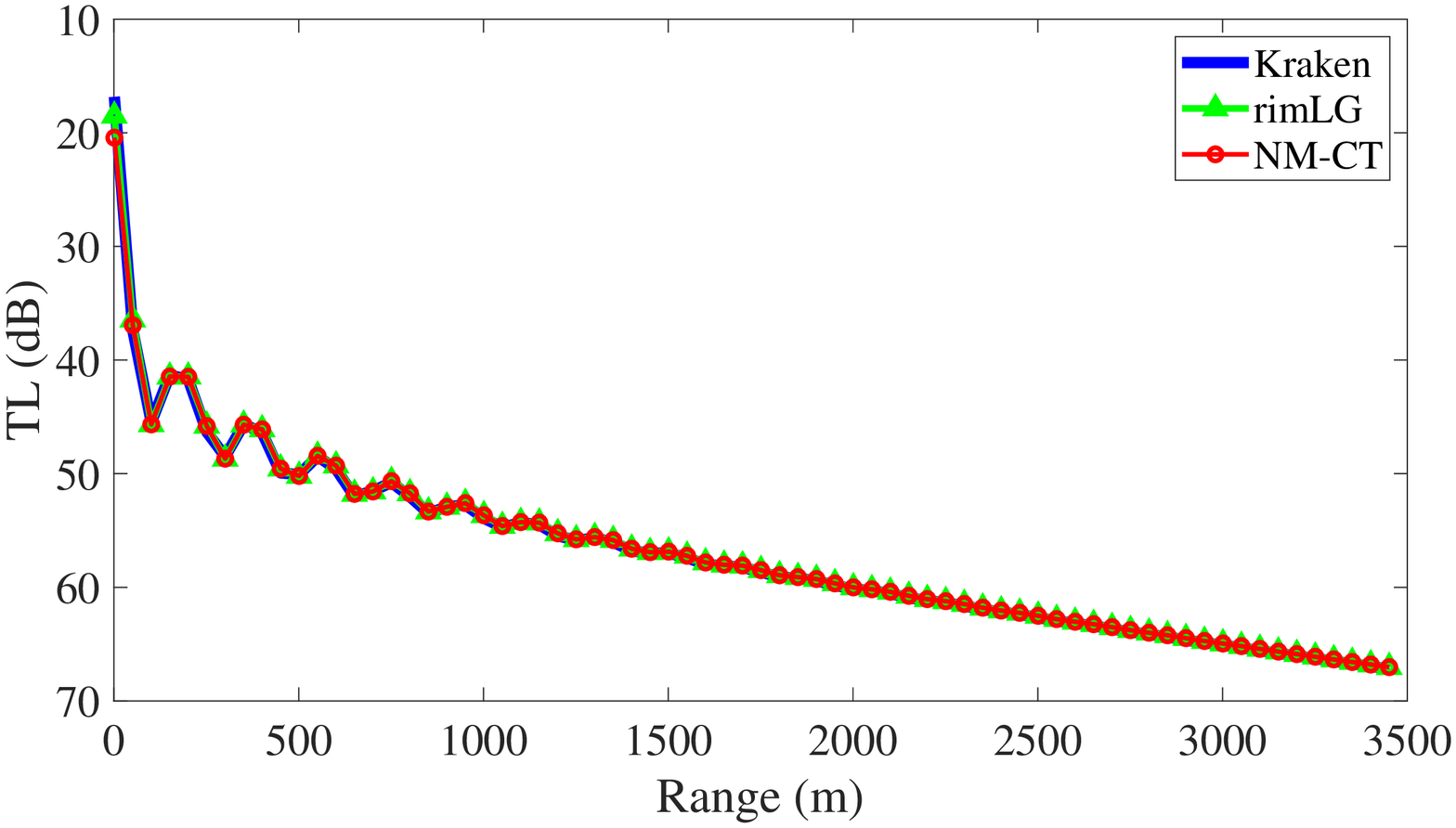}}
\caption{Full-field TLs calculated by Kraken (a), rimLG (b) and NM-CT (c); the black dotted line indicates the interface; TL curves at a depth of 10 m calculated by the three programs in example 2 ($f$=20 Hz).}
\label{fig:4}
\end{figure}

Example 2 is the simplest layered waveguide with attenuated sediment, in which the sound speed, density and attenuation are uniform constants. Table \ref{tab3} shows the horizontal wavenumbers calculated by the Kraken, KrakenC, rimLG and NM-CT programs at frequencies of 20 Hz and 50 Hz, respectively. Kraken and KrakenC use 1000 discrete points in the vertical direction, and the orders of NM-CT and rimLG are only 20 and 40 at 20 Hz and 50 Hz, respectively. It can be seen in the table that the wavenumbers calculated by the rimLG and NM-CT programs are completely consistent at the two source frequencies, and the wavenumbers calculated by the four programs are basically similar. Fig.~\ref{fig:4}(a)-(c) shows the full-field TLs calculated by Kraken, rimLG and NM-CT at 20 Hz. Fig.~\ref{fig:4}(d) shows the TLs at a depth of 10 m calculated by the three programs. The TL field calculated by NM-CT is almost identical to Kraken. The difference between the TLs obtained from Kraken and NM-CT has been found to be indistinguishable at this plotting accuracy. Table \ref{tab4} lists the running times of the three programs in this example. The time listed in the table is the average of 10 test times. In the test, a single thread of a single process is run on MATLAB 2019a on a HUAWEI MatebookX PRO laptop computer with an Intel i7-8550U processor, and the Fortran compiler used in the test is GFortran 7.4.0. It can be seen in the table that the running time of NM-CT is almost equal to Kraken but much shorter than rimLG. Although they both belong to the class of spectral methods, our NM-CT program is faster than the rimLG program. As one of the possible reasons, rimLG consumes too much time in computing a complex symmetric matrix and a positive definite symmetric matrix to form the generalized eigenvalue problem (see Eq. (4) in \cite{Evans2016}), while our NM-CT program aims to solve a simpler eigenvalue problem shown by Eq.~\eqref{eq:38}.

\begin{table}[htbp]
\scriptsize
\centering
\begin{tabular}{llllll}
\hline
Freq.&
$m$&
Kraken&
KrakenC&
rimLG& 
NM-CT \\
\hline
$f$=20 Hz
&1  &   0.07350 70395+ 
    &   0.07349 26194+
    &   0.07350 28581+ 
    &   0.07350 28581+\\
&   &   0.37693 38382e-3i
    &   0.37466 68873e-3i
    &   0.37597 26294e-3i
    &   0.37597 26294e-3i\\

&2  &   0.04033 76782+ 
    &   0.04034 46000+
    &   0.04040 98897+
    &   0.04040 98898+\\
&   &   0.23700 44780e-2i
    &   0.23763 40719e-2i
    &   0.23757 23752e-2i
    &   0.23757 23752e-2i\\
\hline
$f$=50 Hz
&1&0.20330 26836+
  &0.20329 22496+
  &0.20329 61543+ 
  &0.20329 61543+\\
& &0.14684 64650e-3i
  &0.14474 24293e-3i
  &0.14552 80250e-3i
  &0.14552 80250e-3i\\

&2&0.18325 56952+ 
  &0.18318 27296+
  &0.18320 16596+ 
  &0.18320 16596+\\
& &0.72305 34643e-3i
  &0.71311 18231e-3i
  &0.71805 23083e-3i
  &0.71805 23083e-3i\\

&3&0.16334 43629+ 
  &0.16336 23823+
  &0.16348 65836+ 
  &0.16348 65836+\\
& &0.44671 05031e-2i
  &0.44867 81267e-2i
  &0.44892 27771e-2i
  &0.44892 27771e-2i\\

&4&0.14201 67602+
  &0.14188 87957+
  &0.14195 94443+ 
  &0.14195 94443+\\
& &0.25964 36347e-2i
  &0.26003 91729e-2i
  &0.26101 78399e-2i
  &0.26101 78399e-2i\\

&5&0.11339 56065+ 
  &0.11358 52045+
  &0.11371 57329+ 
  &0.11371 57329+\\
& &0.47551 81270e-2i
  &0.47349 61331e-2i
  &0.47261 24780e-2i
  &0.47261 24780e-2i\\
\hline
\end{tabular}
\caption{\label{tab3}Comparison of $k_{r,m}$ of example 2.}
\end{table}

\begin{table}[htbp]
\footnotesize
\centering
\begin{tabular}{ccccc}
\hline
Freq.&
Kraken&
NM-CT(Fortran)&
rimLG&
NM-CT(MATLAB)\\
\hline
$f$=20 Hz& 0.29 s& 0.22 s& 15.16 s& 0.12 s \\
$f$=50 Hz& 0.35 s& 0.24 s& 15.42 s & 0.16 s \\
\hline
\end{tabular}
\caption{\label{tab4}Comparison of the running time of example 2.}
\end{table}

\begin{figure}[htbp]
\centering
\includegraphics[width=\linewidth] {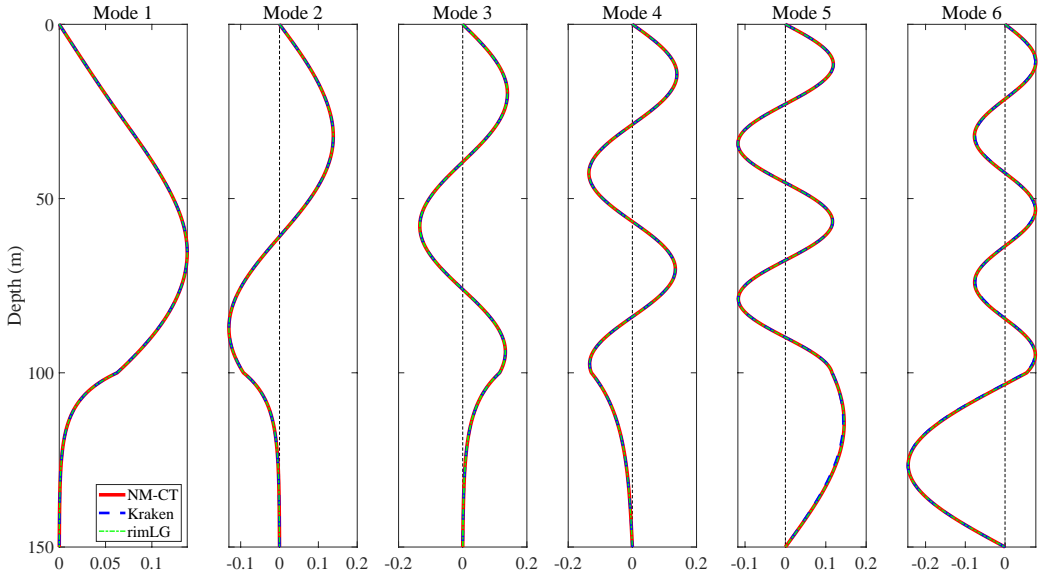}
\caption{The shapes of the first six modes of example 3, were calculated by Kraken (blue dashed line), rimLG (green dashed line) and NM-CT (red solid line) at 50 Hz.}
\label{fig:5}
\end{figure}

In example 3, a surface duct sound speed profile is used in the water column. This sound speed profile is not smooth enough; thus, it requires more spectral truncation order. In this example, there is no attenuation in the bottom sediment, so the results of KrakenC are not shown. Table \ref{tab5} shows the horizontal wavenumbers calculated by the Kraken, rimLG and NM-CT programs at frequencies of 50 Hz and 100 Hz. The Kraken program has 1500 discrete points in the vertical direction, and the orders of NM-CT and rimLG are $N$=500 at 50 Hz and $N$=1000 at 100 Hz, respectively. NM-CT matches Kraken with five and six significant digits for $f$ = 50 Hz and 100 Hz, respectively. Fig.~\ref{fig:5} shows the six modes calculated by NM-CT at 50 Hz; in Fig.~\ref{fig:5}, the modes obtained from Kraken and rimLG are also shown. The difference between the mode shapes obtained from the three programs is found to be indistinguishable for this plotting accuracy. This finding is true for all problems without bottom attenuation. This result shows that the modes calculated by NM-CT are in good agreement with the Kraken results. Table \ref{tab6} lists the running times of the three programs in this example. The conclusion is the same as in example 1 and example 2; NM-CT is much faster than rimLG and slightly slower than Kraken.

\begin{table}[htbp]
\scriptsize
\centering
\begin{tabular}{ccccc}
\hline
Freq.&
$m$&
Kraken&
rimLG&
NM-CT\\
\hline
$f$=50Hz
    &1 &0.20456 66998 &0.20456 67640 & 0.20456 67092\\
    &2 &0.19825 45322 &0.19825 46184 & 0.19825 45356\\
    &3 &0.18790 98058 &0.18790 99137 & 0.18790 98062\\
    &4 &0.17209 24614 &0.17209 26321 & 0.17209 24594\\
    &5 &0.15078 56837 &0.15078 71664 & 0.15078 56682\\
    &6 &1418862 39679 &0.14189 02497 & 0.14188 62352\\
\hline
$f$=100Hz
    &1  &0.41201 50812  &0.41201 51634  &0.41201 50943\\
    &2  &0.40748 98115  &0.40748 99355  &0.40748 98294\\
    &3  &0.40194 64244  &0.40194 65268  &0.40194 64373\\
    &6  &0.37130 65141  &0.37130 66596  &0.37130 65199\\
    &7  &0.35529 03522  &0.35529 05021  &0.35529 03512\\
    &8  &0.33589 86072  &0.33589 88120  &0.33589 86089\\
    &11 &0.29020 30496  &0.29020 93565  &0.29020 30549\\
    &12 &0.28087 06224  &0.28087 37829  &0.28087 06077\\
    &13 &0.25576 83615  &0.25577 45190  &0.25576 84904\\
\hline
\end{tabular}
\caption{\label{tab5}Comparison of $k_{r,m}$ of example 3.}
\end{table}

\begin{table}[htbp]
\footnotesize
\centering
\begin{tabular}{ccccc}
\hline
Freq.&
Kraken&
NM-CT(Fortran)&
rimLG&
NM-CT(MATLAB) \\
\hline
$f$=50 Hz& 0.31 s& 8.37 s& 1348.85 s& 4.23 s\\
$f$=100 Hz& 0.36 s& 46.76 s& 63826.37 s& 51.42 s\\
\hline
\end{tabular}
\caption{\label{tab6}Comparison of the running time of example 3.}
\end{table}

Example 4 adds bottom attenuation 0.5 dB/$\lambda$ based on example 3. Table \ref{tab7} shows the horizontal wavenumbers calculated by the Kraken, KrakenC, rimLG and NM-CT programs at frequencies of 50 Hz and 100 Hz. It can be seen in the table that the wavenumbers calculated by the four programs are still similar, which shows that NM-CT can obtain credible results regardless of whether the bottom is attenuated.
\begin{table}[htbp]
\scriptsize
\centering
\begin{tabular}{llllll}
\hline
Freq.&
$m$&
Kraken&
KrakenC&
rimLG& 
NM-CT\\
\hline
$f$=50 Hz
&1  &   0.20456 65851+ 
    &   0.20456 66180+
    &   0.20456 65949+     
    &   0.20456 65399+\\        
&   &   0.84774 86844e-5i
    &   0.83525 56810e-5i
    &   0.83563 28501e-5i 
    &   0.83558 18884e-5i\\  

&2  &   0.19825 42502+  
    &   0.19825 39711+
    &   0.19825 42260+
    &   0.19825 41431+\\
&   &   0.20928 28166e-4i
    &   0.20592 49724e-4i
    &   0.20601 43841e-4i
    &   0.20599 96787e-4i\\
    
&3  &   0.18790 92721+  
    &   0.18790 89290+
    &   0.18790 91494+ 
    &   0.18790 90428+\\
&   &   0.39680 46830e-4i
    &   0.39122 93174e-4i
    &   0.39142 56820e-4i
    &   0.39139 51987e-4i\\
    
&4  &   0.17209 14240+ 
    &   0.17208 97671+
    &   0.17209 05414+
    &   0.17209 03689+\\
&   &   0.77265 86281e-4i
    &   0.76662 97284e-4i
    &   0.76718 40570e-4i
    &   0.76707 12343e-4i\\
    
&5  &   0.15077 98663+ 
    &   0.15073 87951+
    &   0.15074 45128+
    &   0.15074 30467+\\
&   &   0.47798 91588e-3i
    &   0.47546 02209e-3i
    &   0.47667 11983e-3i
    &   0.47635 77140e-3i\\
    
&6  &   0.14187 53523+ 
    &   0.14191 73773+
    &   0.14193 12614+
    &   0.14192 72116+\\
&   &   0.11089 56529e-2i
    &   0.11106 71974e-2i
    &   0.11096 78864e-2i
    &   0.11099 31242e-2i\\
\hline
$f$=100 Hz
&1  &   0.41201 50087+  
    &   0.41201 49345+
    &   0.41201 50486+
    &   0.41201 49733+\\        
&   &   0.53367 85874e-5i
    &   0.52718 84909e-5i
    &   0.52746 05158e-5i
    &   0.52744 25387e-5i\\  

&2  &   0.40748 96727+ 
    &   0.40748 95194+
    &   0.40748 97232+
    &   0.40748 96199+\\
&   &   0.10237 50490e-4i
    &   0.10098 91480e-4i
    &   0.10104 23904e-4i
    &   0.10103 67151e-4i\\
    
&3  &   0.40194 62152+
    &   0.40194 59869+
    &   0.40194 62242+
    &   0.40194 61277+\\
&   &   0.15473 27718e-4i
    &   0.15240 03585e-4i
    &   0.15247 38530e-4i
    &   0.15246 90493e-4i\\
    
&6  &   0.37130 58929+
    &   0.37130 51598+
    &   0.37130 57092+
    &   0.37130 55634+\\
&   &   0.45967 88595e-4i
    &   0.45379 25969e-4i
    &   0.45402 98311e-4i
    &   0.45400 48023e-4i\\
    
&7  &   0.35528 94757+
    &   0.35528 81718+
    &   0.35528 89150+
    &   0.35528 87638+\\
&   &   0.64790 73341e-4i
    &   0.64149 75073e-4i
    &   0.64188 59391e-4i
    &   0.64184 34819e-4i\\
    
&8  &   0.33589 72138+
    &   0.33589 37153+
    &   0.33589 48592+
    &   0.33589 46561+\\
&   &   0.10299 38733e-3i
    &   0.10229 27681e-3i
    &   0.10238 88776e-3i
    &   0.10237 32380e-3i\\
    
&11 &   0.29018 11398+
    &   0.29003 57030+
    &   0.29006 30258+
    &   0.29005 65311+\\
&   &   0.22085 62615e-2i 
    &   0.22281 09393e-2i
    &   0.22319 59465e-2i
    &   0.22309 85787e-2i\\
    
&12 &   0.28085 83769+
    &   0.28100 43200+
    &   0.28101 60710+
    &   0.28101 30935+\\
&   &   0.10257 56310e-2i
    &   0.10047 80875e-2i
    &   0.10013 84950e-2i
    &   0.10022 50558e-2i\\
    
&13 &   0.25574 78256+
    &   0.25562 82278+
    &   0.25565 45944+
    &   0.25564 84910+\\
&   &   0.21995 85067e-2i
    &   0.21927 77341e-2i
    &   0.21959 49109e-2i
    &   0.21951 82731e-2i\\
\hline
\end{tabular}
\caption{\label{tab7}Comparison of $k_{r,m}$ of example 4.}
\end{table}

\begin{figure}[htbp]
\centering
\subfigure[]{\includegraphics[width=6.5cm]{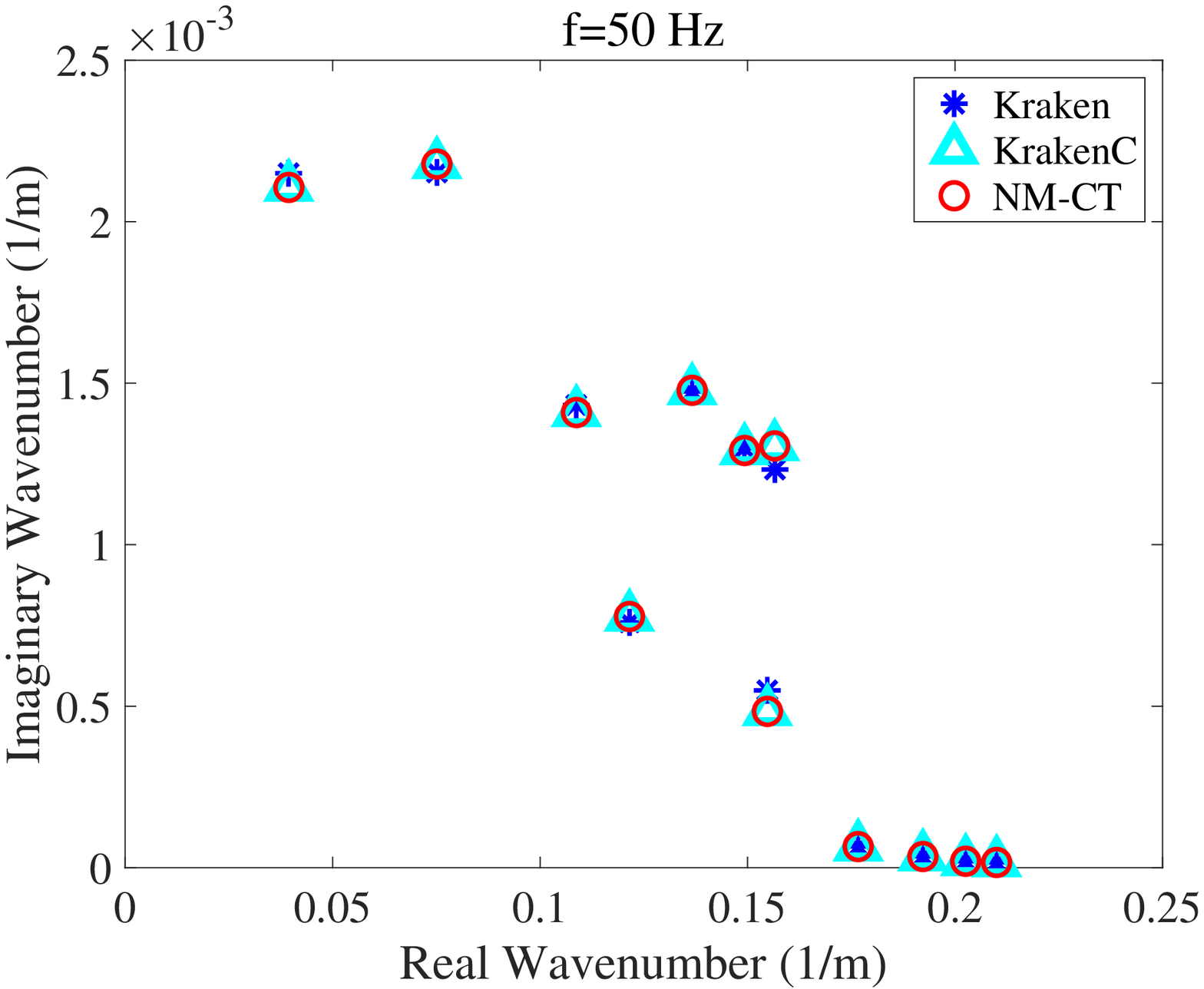}}
\subfigure[]{\includegraphics[width=6.5cm]{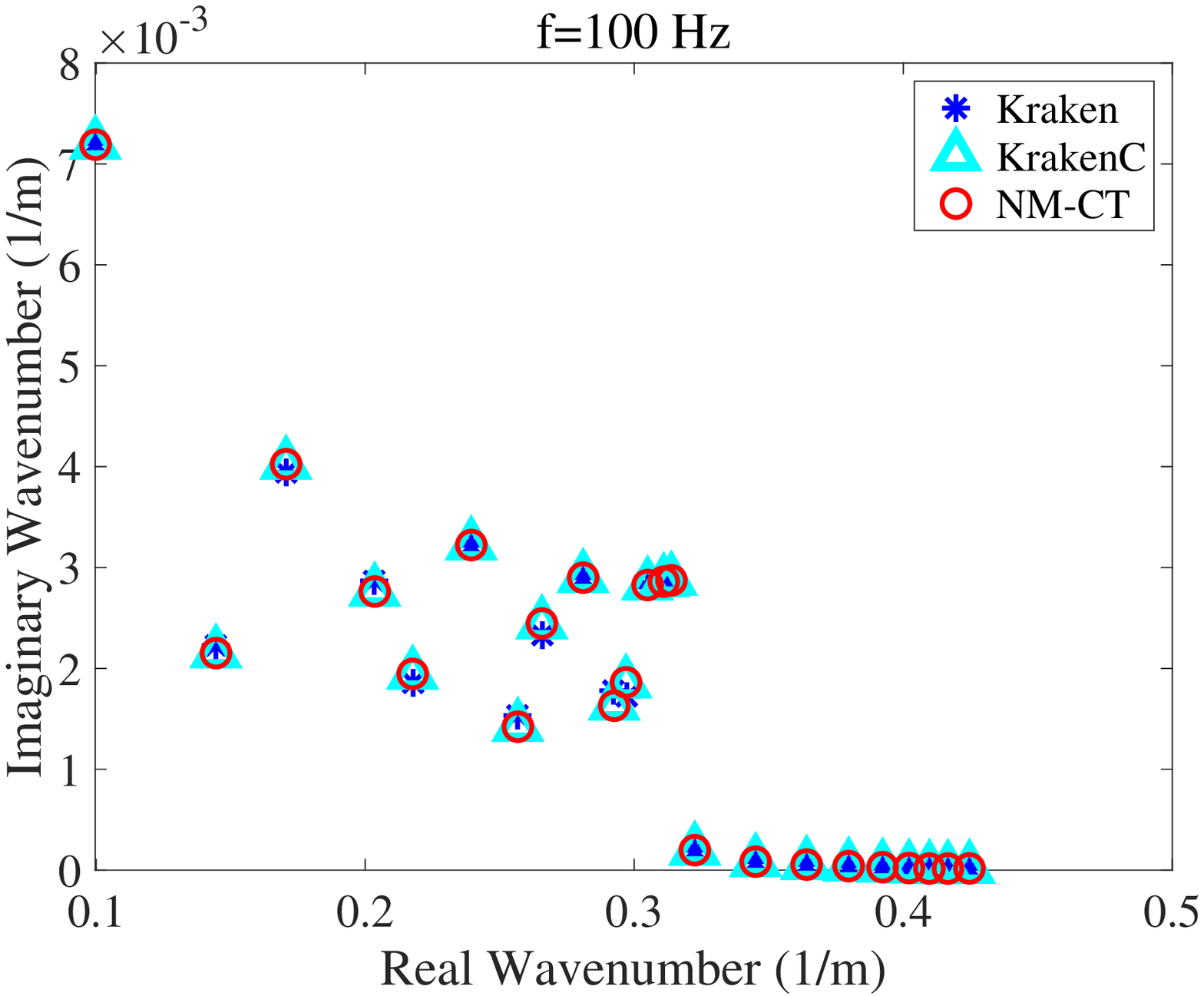}}
\caption{Horizontal wavenumbers $k_r$ calculated by Kraken and NM-CT at 50 Hz (a) and 100 Hz (b) of example 5.}
\label{fig:6}
\end{figure}

\begin{figure}[htbp]
\centering
\includegraphics[width=1\linewidth] {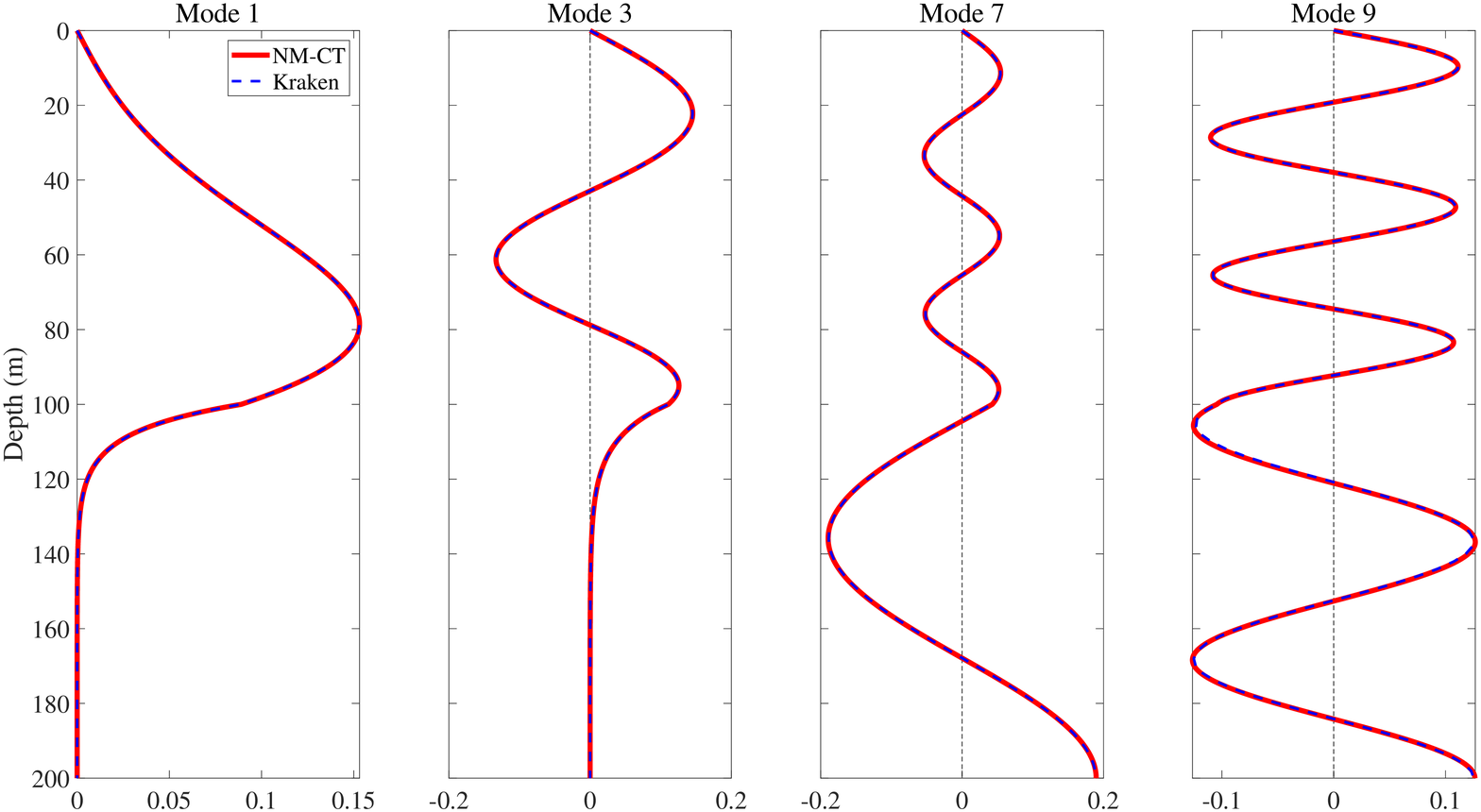}
\caption{Mode shapes for example 5 calculated by Kraken (blue dashed line) and NM-CT (red solid line) at 50 Hz.}
\label{fig:7}
\end{figure}

In example 5, a perfectly rigid boundary condition is used, and a pseudolinear profile and exponentially increasing sound profile are used in the water column and bottom sediment, respectively. Kraken uses 2000 discrete points in the vertical direction, and the orders of NM-CT are $N$=50 and $N$=100 at 50 Hz and 100 Hz, respectively. To intuitively see the wavenumbers calculated by the three programs, Fig.~\ref{fig:6} shows the wavenumbers in the complex plane. It can be seen in the figure that the wavenumbers calculated by NM-CT are closer to KrakenC than Kraken; in general, the results of the three programs are highly similar in this example. Fig.~\ref{fig:7} shows the random four modes in the example of 50 Hz. From the shapes of the modes, we can see that there are subtle and imperceptible differences in the last modes.

Example 6 is very similar to Example 5, except that the lower boundary condition is changed to perfectly free. Table \ref{tab8} shows the horizontal wavenumbers calculated by the Kraken, rimLG and NM-CT programs at frequencies of 50 Hz and 100 Hz. It can be seen in the table that the wavenumbers calculated by the three programs are still similar, which shows that NM-CT can obtain convincing results regardless of whether the bottom is perfectly rigid or free.

\begin{table}[htbp]
\scriptsize
\centering
\begin{tabular}{llllll}
\hline
    Freq.   &
    $m$     &
    Kraken  &
    KrakenC &
    rimLG   &
    NM-CT   \\
\hline
$f$=50 Hz
    &1  &0.21001 77364+
        &0.21001 75130+
        &0.21001 76897+
        &0.21001 76606+  \\
    &   &0.16131 56264e-4i
        &0.16127 40702e-4i
        &0.16135 42837e-4i
        &0.16134 19472e-4i  \\
    
    &2  &0.20257 20089+
        &0.20257 17620+
        &0.20257 19786+ 
        &0.20257 19436+  \\
    &   &0.19856 05398e-4i
        &0.19851 49397e-4i
        &0.19860 68575e-4i
        &0.19859 33106e-4i  \\

    &3  &0.19226 25150+
        &0.19226 20600+
        &0.19226 24339+
        &0.19226 23760+  \\
    &   &0.34538 89758e-4i
        &0.34529 76202e-4i
        &0.34546 06351e-4i
        &0.34543 83693e-4i  \\   

    &4  &0.17666 02717+
        &0.17665 90008+
        &0.17665 96941+ 
        &0.17665 95967+  \\
    &   &0.65210 59254e-4i
        &0.65171 19663e-4i
        &0.65209 40806e-4i
        &0.65205 17663e-4i  \\

    &5  &0.15567 70622+ 
        &0.15553 29067+
        &0.15553 92553+
        &0.15553 90159+  \\
    &   &0.70465 84115e-3i
        &0.69470 12846e-3i
        &0.69754 71573e-3i
        &0.69755 14104e-3i\\

    &6  &0.15289 03336+
        &0.15302 28270+
        &0.15303 20732+
        &0.15303 20250+  \\
    &   &0.97022 58068e-3i
        &0.97893 20248e-3i
        &0.97650 59872e-3i
        &0.97648 88614e-3i  \\

    &7  &0.14377 55359+
        &0.14378 63608+
        &0.14379 95849+
        &0.14379 94008+  \\
    &   &0.14193 21581e-2i
        &0.14205 58800e-2i
        &0.14204 63379e-2i
        &0.14204 72405e-2i  \\

    &8  &0.12835 94066+
        &0.12832 98501+
        &0.12834 25492+
        &0.12834 22181+  \\
    &   &0.13425 76929e-2i
        &0.13453 64528e-2i
        &0.13460 03304e-2i
        &0.13460 02038e-2i\\
\hline
$f$=100 Hz
    &1  &0.42460 86614+ 
        &0.42460 84054+     
        &0.42460 86003+
        &0.42460 85640+  \\ 
    &   &0.17336 02035e-4i 
        &0.17329 20134e-4i
        &0.17338 44497e-4i
        &0.17336 78968e-4i  \\

    &2  &0.41660 99568+
        &0.41660 97405+
        &0.41660 99159+
        &0.41660 98832+  \\
    &   &0.15615 31545e-4i 
        &0.15609 31461e-4i
        &0.15617 30488e-4i
        &0.15615 89033e-4i  \\

    &3  &0.40973 02081+ 
        &0.40972 99945+
        &0.40973 01752+
        &0.40973 01418+  \\
    &   &0.16097 66605e-4i
        &0.16091 58499e-4i
        &0.16099 59230e-4i
        &0.16098 18988e-4i  \\
    
    &7  &0.36408 99387+
        &0.36408 89665+
        &0.36408 95771+
        &0.36408 94714+   \\
    &   &0.55276 64987e-4i
        &0.55247 10681e-4i
        &0.55279 06003e-4i
        &0.55274 10256e-4i  \\

    &8  &0.34512 12700+
        &0.34511 90586+
        &0.34511 99730+
        &0.34511 98231+  \\
    &   &0.83894 99567e-4i 
        &0.83799 71712e-4i
        &0.83862 68939e-4i
        &0.83854 02070e-4i  \\

    &9  &0.32253 76329+
        &0.32251 97554+
        &0.32252 17833+
        &0.32252 15118+  \\
    &   &0.20056 89504e-3i
        &0.19780 03179e-3i
        &0.19818 31732e-3i
        &0.19815 01265e-3i  \\

    &13 &0.29477 93267+
        &0.29474 78357+
        &0.29475 63509+
        &0.29475 60875+  \\
    &   &0.10859 36417e-2i
        &0.10237 39337e-2i
        &0.10246 11119e-2i
        &0.10245 10168e-2i  \\

    &14 &0.28749 10525+
        &0.28761 77672+
        &0.28764 27674+ 
        &0.28764 23007+\\
    &   &0.25902 70973e-2i
        &0.26176 44484e-2i
        &0.26155 55357e-2i
        &0.26156 25849e-2i\\

    &15 &0.27344 09188+ 
        &0.27341 42528+
        &0.27344 18466+ 
        &0.27344 11802+\\
    &   &0.29129 68544e-2i
        &0.29222 85503e-2i
        &0.29229 56121e-2i
        &0.29229 64710e-2i\\
\hline
\end{tabular}
\caption{\label{tab8}Comparison of $k_{r,m}$ of example 6.}
\end{table}

\begin{figure}[htbp]
\centering
\subfigure[]{\includegraphics[width=6.5cm]{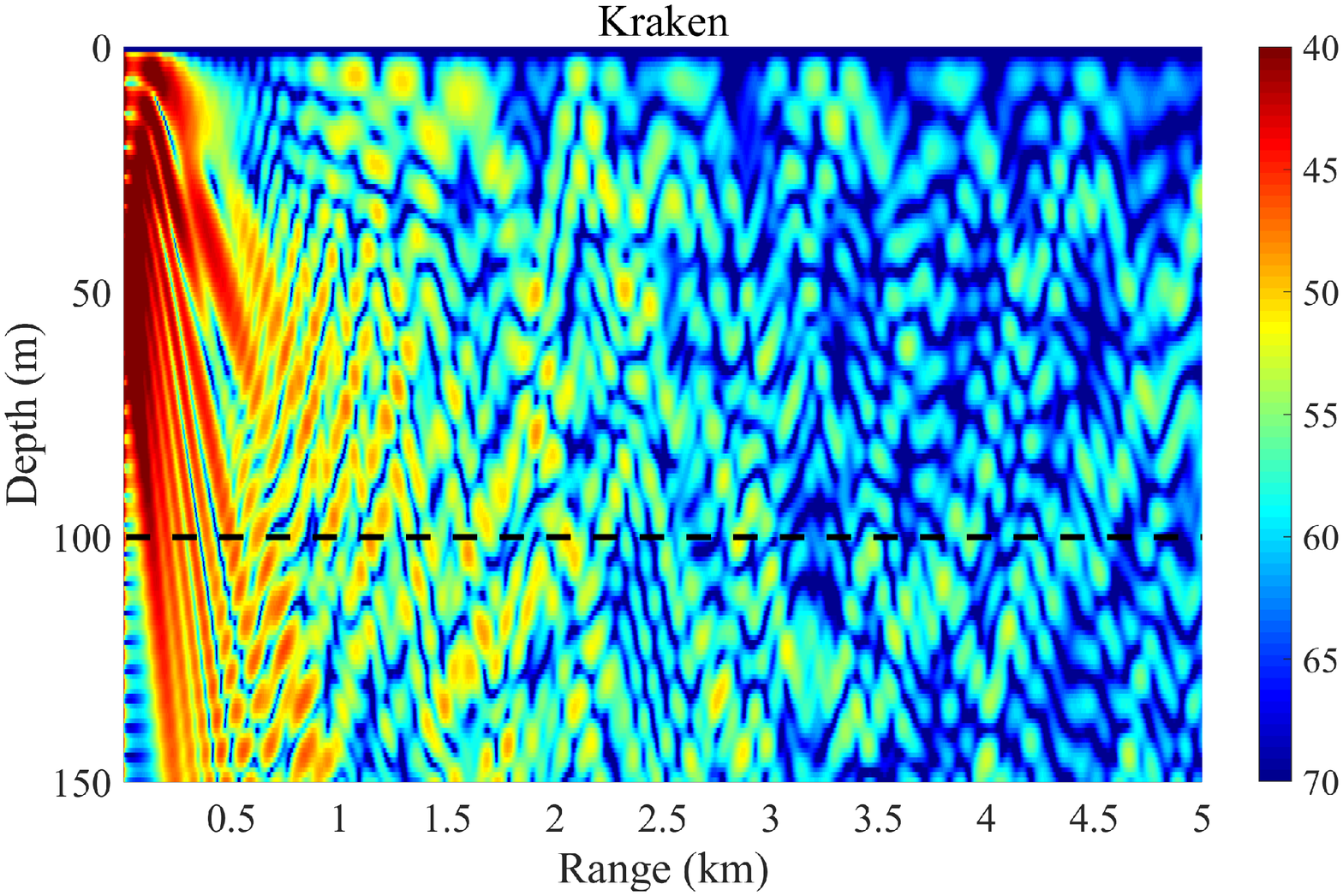}}
\subfigure[]{\includegraphics[width=6.5cm]{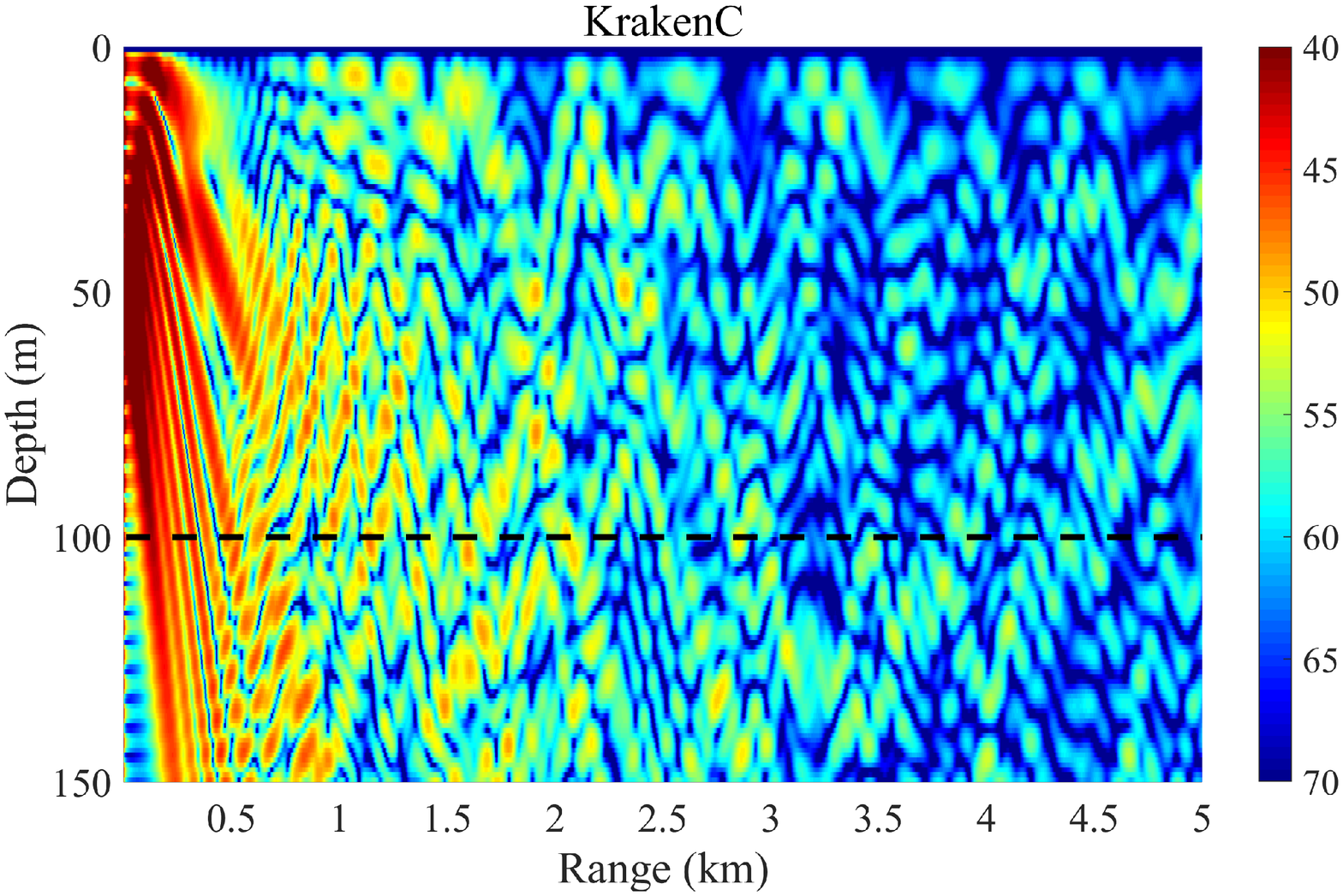}}
\subfigure[]{\includegraphics[width=6.5cm]{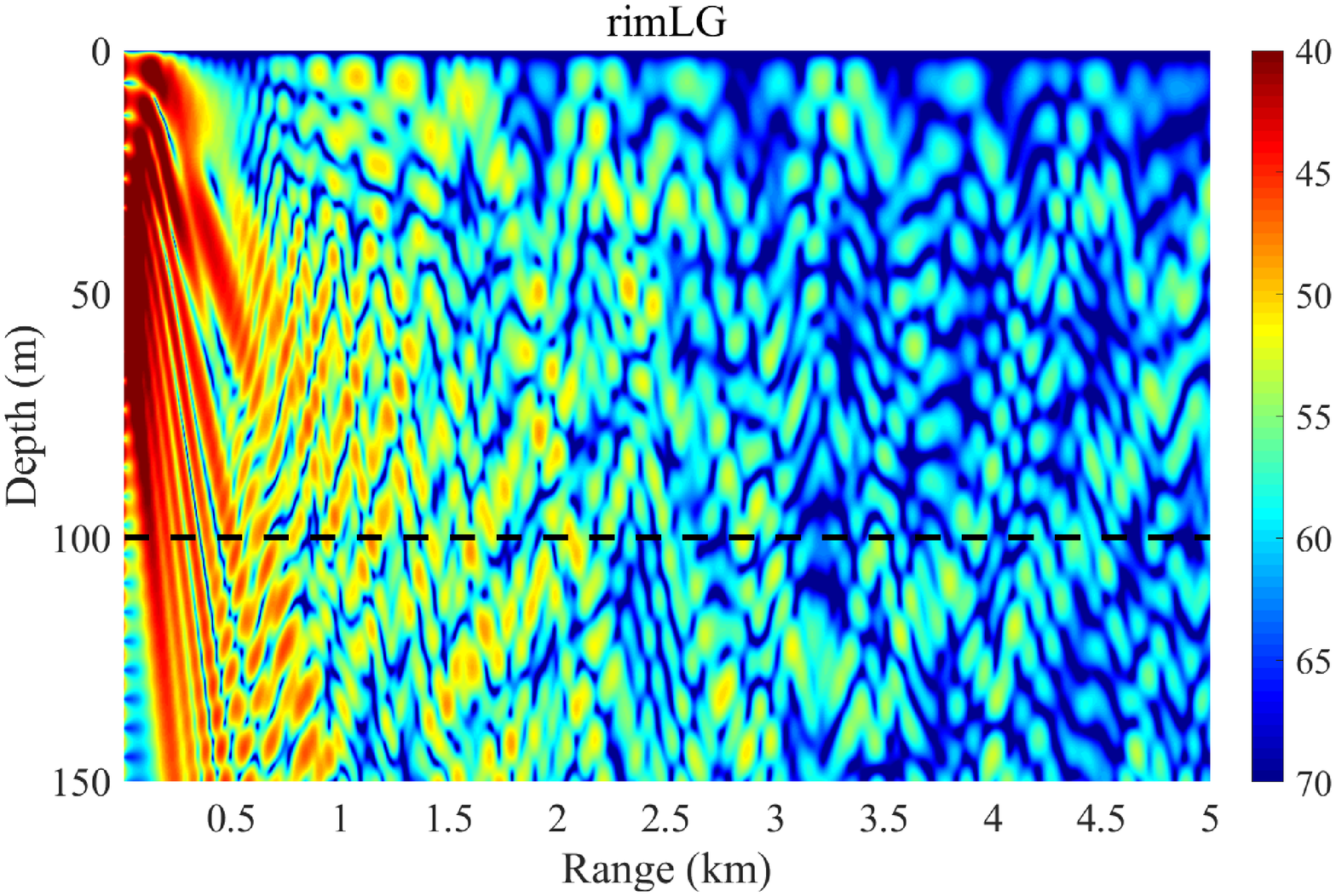}}
\subfigure[]{\includegraphics[width=6.5cm]{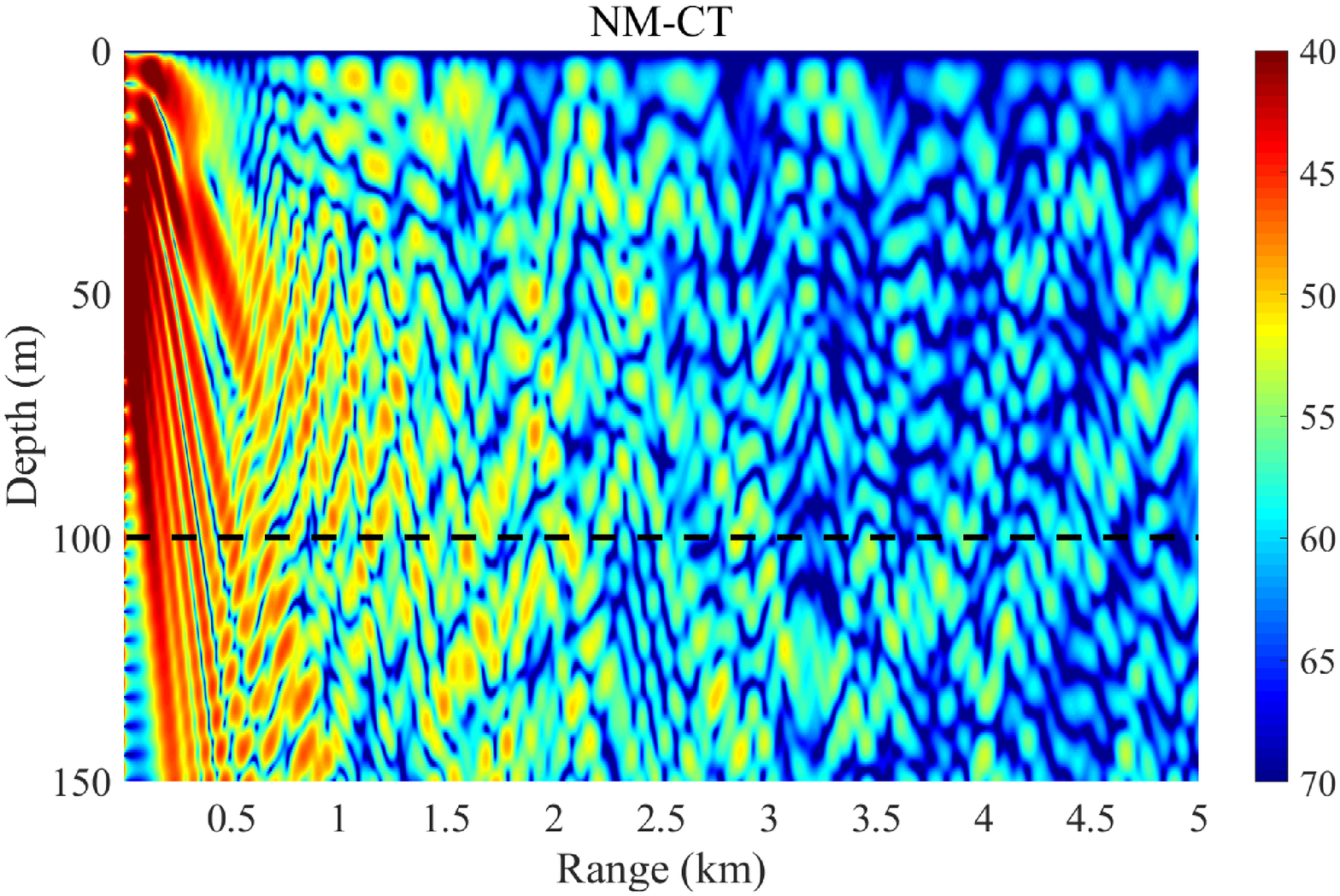}}
\caption{Full-field TLs calculated by Kraken (a), Kraken (b), rimLG (c) and NM-CT (d) of example 7; the black dotted line indicates the interface.}
\label{fig:8}
\end{figure}

Example 7 is an example with the discontinuous second derivative of the sound speed profile from the literature \cite{Evans2016}, which is used only to test the robustness of NM-CT in calculating the sound speed profile with a discontinuous derivative. This is especially important in cases when a large number of measured sound speeds are used. In this example, $f$=250 Hz, the number of discrete points used in Kraken and KrakenC are automatically chosen by the code, and the orders of both rimLG and NM-CT are $N$=500. Table \ref{tab9} shows the horizontal wavenumbers, and Fig. \ref{fig:8} shows the TL fields calculated by the four programs. From the table and figure, we can draw the conclusion that NM-CT can still obtain satisfactory results, which means that although the Chebyshev-Tau method proposed in this article theoretically requires the acoustic profiles to be sufficiently smooth, for the measured data that are not so smooth, NM-CT can also obtain valuable results for reference by using a large $N$.

\begin{table}[htbp]
\scriptsize
\centering
\begin{tabular}{lllll}
\hline
    $m$     &
    Kraken  &
    KrakenC &
    rimLG   &
    NM-CT   \\
\hline
    1   &1.06120 73981+ 
        &1.06120 73980+
        &1.06120 74892+
        &1.06120 73934+   \\
        &0.31452 71656e-7i
        &0.31343 84747e-7i
        &0.31311 49600e-7i
        &0.31309 30200e-7i  \\
    
    4   &1.05908 53065+
        &1.05908 53030+
        &1.05908 68080+ 
        &1.05908 52520+ \\
        &0.49312 57091e-6i
        &0.49134 81574e-6i
        &0.49082 92710e-6i
        &0.49079 67580e-6i   \\    
    
    8   &1.05230 17898+
        &1.05230 17200+
        &1.05230 80953+ 
        &1.05230 15655+  \\
        &0.18612 79366e-5i
        &0.18540 63045e-5i
        &0.18519 88927e-5i
        &0.18518 55544e-5i\\

    16  &1.02726 88405+
        &1.02726 86270+
        &1.02727 57016+
        &1.02726 88998+  \\
        &0.46107 53397e-5i
        &0.45934 31907e-5i
        &0.46093 56507e-5i
        &0.45871 06521e-5i\\   

    24  &0.99057 13001+
        &0.99057 08911+
        &0.99058 83179+
        &0.99057 09466+  \\
        &0.18246 84646e-4i
        &0.18275 41277e-4i 
        &0.18588 97603e-4i
        &0.18221 72908e-4i  \\

    40  &0.93895 77115+  
        &0.93895 69464+     
        &0.94402 33694+
        &0.93895 88486+  \\
        &0.11022 01392e-2i
        &0.11109 67196e-2i 
        &0.82054 60202e-3i
        &0.10524 36440e-2i  \\

    48  &0.92154 89408+
        &0.92153 90152+ 
        &0.92896 10252+
        &0.92154 23293+  \\
        &0.12494 56899e-2i
        &0.12630 46083e-2i
        &0.10808 99895e-2i
        &0.11858 48928e-2i  \\

    50  &0.91755 92536+ 
        &0.91755 79958+  
        &0.92582 79956+
        &0.91755 89811+   \\
        &0.88280 41759e-3i 
        &0.82583 86769e-3i
        &0.82816 76867e-3i
        &0.78282 21102e-3i\\

\hline
\end{tabular}
\caption{\label{tab9}Comparison of $k_{r,m}$ of example 7.}
\end{table}

From the analysis of the above examples, we can see that in general, NM-CT uses fewer discrete points to achieve or exceed Kraken's accuracy, especially when the acoustical profiles are smooth. The small number of discrete points in the vertical direction means that the order of spectral truncation is relatively low. As a result, the calculation amount of the spectral method is small, and the computational speed is not much longer than Kraken. When the acoustical profiles are not smooth enough, the spectral methods (NM-CT and rimLG) require a higher truncation order, which causes an increase in the computational time. For the case of large $N$, the discrete Chebyshev forward transform Eq.~\eqref{eq:17} and inverse transform Eq.~\eqref{eq:18} can be accelerated in $(2.5N \log_2{N}+4N)$ float operations by using the FFT (fast Fourier transforms) shown in Appendix B of \cite{Canuto2006}. In general, compared with Kraken, the NM-CT program has a slower computational speed but higher accuracy. Compared with rimLG, although they are both spectral methods, the computational speed of NM-CT is much faster than that of rimLG when the accuracy is equivalent. To make the comparison more convenient, NM-CT uses the same truncation orders in both the water column and bottom sediment. In fact, we have tried to use different $N$ in the water column and bottom sediment; in actual operation, the truncation orders in the two layers can be freely specified in the input file of the NM-CT according to the complexity of the profiles of the two layers. The results show that NM-CT can still obtain sufficiently accurate horizontal wavenumbers and sound fields in this configuration.

\section{Conclusion and future work}

The Chebyshev-Tau spectral method provides a simple matrix procedure for finding two-layer underwater acoustic normal modes when the sea surface, bottom and interface are range independent. This method first interpolates the acquired data of the sound speed, density and attenuation profiles to the Gauss-Lobatto points.
After a permutation on the rows and columns in the coefficient matrix, a complex eigensystem for solving wavenumbers and modal spectral coefficients is formed that can be solved by applying numerical libraries and algorithms.
The validity of this Chebyshev-Tau spectral method (NM-CT program) is demonstrated in comparison with the finite difference method (Kraken program and its variations) and Legendre-Galerkin spectral method (rimLG program).
From the perspective of the computational speed, NM-CT is superior to rimLG. From the perspective of computational accuracy, NM-CT is also sufficiently high compared with Kraken and rimLG.
In fact, this method can also address more complex situations, such as irregular changes in the density and attenuation coefficient with depth, but real ocean data are too difficult to obtain. Therefore, this article does not show particularly complicated numerical examples. Notably, in the case where the sound speed, density and attenuation profiles are not smooth enough, the Chebyshev-Tau spectral method can use a higher truncation order to obtain convincing results.

This article simply divides the actual ocean into two layers, water and sediment. In the actual ocean, the sediment may be composed of silt or sand with completely different densities and attenuations. In this case, the sediment should be divided into more intermittent layers. It can be seen from the derivation in Sect.~\ref{sec:3} that our method has the potential to work in multilayered environments. Regarding the issue that the computational time is longer than Kraken, we will try to speed up the running time through parallel computing technology in the next work.

\section*{Acknowledgments}

This work was supported by the National Key Research and Development Program of China [grant number 2016YFC1401800]; the National Natural Science Foundation of China [grant numbers 61972406, 51709267]; and the ``Double First-Class" Construction Guidance Project of National University of Defense Technology [grant number 4345161111L].

\bibliographystyle{elsarticle-num}
\bibliography{sample}
\end{document}